\pdfoutput=1 
\documentclass[12pt]{article}
\usepackage[english]{babel}
\usepackage{hyperref}
\usepackage{xcolor}
\usepackage{pslatex}
\usepackage{amsmath}
\usepackage{amssymb}
\usepackage{bbm}
\usepackage{euscript}
\usepackage{epsfig}
\usepackage{graphicx}
\usepackage{setspace}
\usepackage{cite}

\newcommand{\mincas}{{\tt MINCAS}}
\newcommand{\foam}{{\tt FOAM}}

\begin{document}
\begin{titlepage}

\begin{center}
\end{center}
\begin{flushright}
\bf IFJPAN-IV-2018-19 \\
    November 2018
\end{flushright}

\vspace{5mm}
\begin{center}
    {\LARGE\bf 
       Solutions of evolution equations \vspace{2mm}\\ 
       for medium-induced QCD cascades$^{\star}$} 
\end{center}

\vskip 5mm
\begin{center}
{\large K.\ Kutak$^a$,
        W.\ P\l{}aczek$^b$ and
        R.\ Straka$^c$
        }
\\
\vskip 2mm
{\em $^a$Institute of Nuclear Physics, Polish Academy of Sciences,\\
  ul.\ Radzikowskiego 152, 31-342 Krak\'ow, Poland}
\\
\vspace{1mm}
{\em $^b$Marian Smoluchowski Institute of Physics, Jagiellonian University,\\
ul.\ \L{}ojasiewicza 11, 30-348 Krak\'ow, Poland}
\\
\vspace{1mm}
{\em $^c$AGH University of Science and Technology, Krak\'ow, Poland}
\end{center}
 
\vspace{20mm}
\begin{abstract}
\noindent
In this paper we present solutions of evolution equations for inclusive distribution 
of gluons as produced by jet traversing quark--gluon plasma. 
We reformulate the original equations in such a form that
virtual and unresolved-real emissions as well as unresolved collisions with medium 
are resummed in a Sudakov-type form factor. The resulting integral equations are then solved
most efficiently with use of newly developed Markov Chain Monte Carlo
algorithms implemented in a dedicated program called \mincas.
Their results for a gluon energy density are compared with an analytical
solution and a differential numerical method. 
Some results for gluon transverse-momentum distributions are also presented.
They exhibit interesting patterns not discussed so far in the literature,
in particular a departure from the Gaussian behaviour --
which does not happen in approximate analytical solutions.
\end{abstract}

\vspace{40mm}
\footnoterule
\noindent
{\footnotesize
$^{\star}$This work is partly supported by
 the Polish National Science Centre grant no. DEC-2017/27/B/ST2/01985.
}

\end{titlepage}

\section{Introduction}
\label{sec:Intro}

Quantum chromodynamics (QCD) is the well established theory of strong interactions. However, there are QCD phenomena that still require better understanding. One of such phenomena is jet quenching predicted in \cite{Gyulassy:1990ye,Wang:1991xy} and already observed in the context of the RHIC physics\cite{Adler:2002tq} 
(for an overview see \cite{Andrews:2018jcm,Wiedemann:2009sh} and references therein), 
i.e. stopping  of a hadronic jet produced in an early stage of heavy ion collisions and propagating through quark-gluon
plasma (QGP) which is formed in a later stage of the collisions. With the LHC being in operation, the jet quenching can be
observed at much higher available energies in collisions of lead nuclei \cite{Aad:2010bu}. 
Still one of the open problems is to understand the details of the jet--QGP interaction mechanism and a pattern of
energy loss. Various approaches have been proposed which
differ in assumptions about properties of plasma and jet--plasma interactions.
Examples are: 
the kinetic theory assuming that the jet--plasma interactions can be described within a weak-coupling regime of QCD \cite{Baier:2000mf,Baier:2000sb,Jeon:2003gi,Zakharov:1996fv,Zakharov:1997uu,Zakharov:1999zk,Baier:1994bd,Baier:1996vi,Arnold:2002ja,Ghiglieri:2015ala}, the AdS/CFT models where one assumes the plasma to be strongly coupled \cite{Liu:2006ug,Chesler:2014jva} or the classical-field-theory-based approach \cite{Carrington:2015xca}
(for reviews see \cite{Baier:2000mf,Mehtar-Tani:2013pia,Ghiglieri:2015zma,Blaizot:2015lma,Busza:2018rrf}). 
Some of the mentioned formalisms are implemented in Monte Carlo event generators
\cite{Salgado:2003gb,Zapp:2008gi,Armesto:2009fj,Schenke:2009gb,Lokhtin:2011qq,Casalderrey-Solana:2014bpa}.

In this paper we look closer at the results obtained in~\cite{Blaizot:2013vha,Blaizot:2014rla} and focus on an analysis of the generation of transverse momenta via cascades of subsequently emitted jets from an energetic jet traversing QGP. 
In this approach the plasma is modelled by static centres and the jet interacts with it weakly. 
Using equations for the energy loss of the jet traversing QGP, the authors of \cite{Blaizot:2013vha,Blaizot:2014rla,Fister:2014zxa} found
the process to have turbulent properties, i.e.\ the energy is transported from large values of $x$ to low values of $x$
without being accumulated at intermediate values. In this paper we investigate a more exclusive equation, i.e.\ 
the equation which describes time evolution of longitudinal as well as transverse momenta distributions of gluons 
emitted from the energetic jet. So far this equation has not been solved numerically, and the analytical as well as numerical analyses 
are limited to some special cases \cite{Blaizot:2014rla,Blaizot:2014ula,Iancu:2015uja} where for instance part the equation leading to broadening of transverse momentum is simplified
or it is included as an input distribution. The simplified analysis suggests that the distribution of gluon
transverse momenta being a solution of the equation remains Gaussian \cite{Blaizot:2014ula}. 
We, however, find that the exact numerical solution of the evolution equation given in 
Ref.~\cite{Blaizot:2014rla} is not Gaussian. 
Furthermore, the numerical method allows to test consequences of the assumptions about properties of the medium for  
distribution patterns of the jets emitted from the hard jet. 

The paper is organised as follows. In Section~2, we introduce and overview the
equations for the jet energy distribution and for the inclusive gluon distribution. In Section~3, we present reformulation of
the above equations making use of Sudakov-like resummation, i.e.\ 
we resum virtual and unresolved-real emissions as well as unresolved collisions with the medium 
of minijets from the highly energetic jet in form of a Sudakov-type form factor. 
Then, we provide formal iterative solutions of these equations. 
In Section~4, we propose a Markov Chain Monte Carlo (MCMC) algorithms 
for numerical solutions of the above equations.
In Section~5, we describe a numerical algorithm for solving the integro-differential 
equation for the jet energy distribution which is based on application of the Runge--Kutta method 
and discuss its limitations in obtaining high accuracy solutions.
In Section~6, first, we present numerical results from the MCMC method for the jet energy distribution
and compare them with an analytical solution as well as with results from the differential
Runge--Kutta-based method.
Then, we show and discuss some results for the jet transverse-momentum distributions obtained with the MCMC method. 
We summarise our work and present its outlook in Section~7.
Finally, in Appendix~A we provide some further details on the MCMC algorithms,
in particular we describe a combination of the branching Monte Carlo method with
the importance sampling.

\section{Evolution equations}
\label{sec:EvolEq}

The evolution equation for gluon transverse-momentum-dependent
distribution $D(x,\mathbf{k},t)$ in the dense medium, 
obtained under the assumption that the momentum transfer in the kernel is small, reads
\cite{Blaizot:2014rla}
\begin{equation}
\begin{aligned}
\frac{\partial}{\partial t} D(x,\mathbf{k},t) = & \: \frac{1}{t^*} \int_0^1 dz\, {\cal K}(z) \left[\frac{1}{z^2}\sqrt{\frac{z}{x}}\, D\left(\frac{x}{z},\frac{\mathbf{k}}{z},t\right)\theta(z-x) 
- \frac{z}{\sqrt{x}}\, D(x,\mathbf{k},t) \right] \\
+& \int \frac{d^2\mathbf{q}}{(2\pi)^2} \,C(\mathbf{q})\, D(x,\mathbf{k}-\mathbf{q},t),
\end{aligned}
\label{eq:ktee1}
\end{equation}
where 
\begin{equation}
{\cal K}(z) = \frac{\left[f(z)\right]^{5/2}}{\left[z(1-z)\right]^{3/2}}, 
\quad   f(z) = 1 - z + z^2, 
\qquad  0 \leq x \leq 1, 
\label{eq:kernel1}
\end{equation}
is the $z$-kernel function, and
\begin{equation}
 \frac{1}{t^*}  = \frac{\bar{\alpha}}{\tau_{\rm br}(E)} = \bar{\alpha}\sqrt{\frac{\hat{q}}{E}}, 
\qquad \bar{\alpha} = \frac{\alpha_s N_c}{\pi},
\label{eq:tstar}
\end{equation}
where $t^*$ is a stopping time, i.e.\ the time at which the energy of an incoming parton has been radiated in form 
of soft gluons, 
$E$ is the energy of the incoming parton, 
$x$ -- its longitudinal momentum fraction, 
$\mathbf{k}=(k_x,k_y)$ -- its transverse-momentum vector,
$\hat{q}$ -- the quenching parameter,
$\alpha_s$ -- the QCD coupling constant
and $N_c$ -- the number of colours.

The collision kernel $C(\mathbf{q})$ is given by
\begin{equation}
C(\mathbf{q}) = w(\mathbf{q}) - \delta(\mathbf{q}) \int d^2\mathbf{q'}\,w(\mathbf{q'})\,,
\label{eq:Cq}
\end{equation}
where the function $w(\mathbf{q})$, 
which models out-of-equilibrium momentum distributions of medium quasi-particles, 
takes the form~\cite{Blaizot:2014rla}
\begin{equation}
 w(\mathbf{q}) = \frac{16\pi^2\alpha_s^2N_cn}{\mathbf{q}^4}\,,
\label{eq:wq1}
\end{equation}
with $\mathbf{q}=(q_x,q_y)$ being transverse-momentum vector 
and $n$ -- the density of scatterers.
However, we can also consider a situation where the quark--gluon plasma equilibrates 
and the above transverse-momentum distribution assumes the form \cite{Gyulassy:1993hr}
\begin{equation}
 w(\mathbf{q}) = \frac{16\pi^2\alpha_s^2N_cn}{\mathbf{q}^2(\mathbf{q}^2+m_D^2)}\,,
\label{eq:wq2}
\end{equation}
where $m_D$ is the Debye mass of the medium quasi-particles.
In the following we shall consider both the above expressions for  $w(\mathbf{q})$.

After integration of Eq.~(\ref{eq:ktee1}) over the transverse
momentum $\mathbf{k}$ one obtains the evolution equation for
gluon energy density \cite{Blaizot:2014rla}
\begin{equation}
\frac{\partial}{\partial t} D(x,t) = \frac{1}{t^*} \int_0^1 dz\, {\cal K}(z) \left[ \sqrt{\frac{z}{x}}\, D\left(\frac{x}{z},t\right)\theta(z-x) - \frac{z}{\sqrt{x}}\, D(x,t) \right]. 
\label{eq:eee1}
\end{equation}
This integro-differential equation can be solved analytically
for a simplified case of $f(z) = 1$ and $D(x,t=0) = \delta(1-x)$
\cite{Blaizot:2014rla}:
\begin{equation}
D(x,\tau) = \frac{\tau}{\sqrt{x}(1-x)^{3/2}}\,\exp\left(-\pi\,\frac{\tau^2}{1-x}\right)\,,
\label{eq:Smplf}
\end{equation}
where $\tau = t/t^*$.

\section{Integral equations and iterative solutions}
\label{sec:IntEqsItSol}

Let us rewrite Eq.~(\ref{eq:ktee1}) by moving all terms with the minus sign from RHS to LHS and using $\tau = t/t^*$:
\begin{equation}
\begin{aligned}
& \frac{\partial}{\partial \tau} D(x,\mathbf{k},\tau) 
 + D(x,\mathbf{k},\tau) \left[ \frac{1}{\sqrt{x}} \int_0^1 dz\, z{\cal K}(z) + t^*  \int d^2 \mathbf{q} \,\frac{w(\mathbf{q})}{(2\pi)^2} \right] \\
& =
 \int_0^1 dz\, {\cal K}(z) \frac{1}{z^2}\sqrt{\frac{z}{x}}\, D\left(\frac{x}{z},\frac{\mathbf{k}}{z},\tau\right)\theta(z-x) 
 + 
t^* \int d^2\mathbf{q} \,\frac{w(\mathbf{q})}{(2\pi)^2}\, D(x,\mathbf{k}-\mathbf{q},\tau)\,.
\end{aligned}
\label{eq:ktee2}
\end{equation}
Then, after introducing the following notation:
\begin{eqnarray}
\Phi(x) &=& 
 \frac{1}{\sqrt{x}}  \int_0^{1-\epsilon} dz\, z {\cal K}(z),
\label{eq:Phi} \\
W &=& 
    t^* \int_{|\mathbf{q} |>q_{\mathrm{min}}} d^2 \mathbf{q}
    \,\frac{w(\mathbf{q})}{(2\pi)^2}, 
\label{eq:W} \\
\Psi(x) &=&\Phi(x) + W\,,
\label{eq:Psi}
\end{eqnarray}
we can write
\begin{equation}
\begin{aligned}
& \frac{\partial}{\partial \tau} D(x,\mathbf{k},\tau) 
 + D(x,\mathbf{k},\tau) \Psi(x) \\
& =
\int_0^1 dy  \int_0^1 dz  \int d^2\mathbf{k}'   \int d^2\mathbf{q} 
\left[ \sqrt{\frac{z}{x}} z{\cal K}(z) \theta(1-\epsilon - z) \delta(\mathbf{q}) 
 + 
t^* \frac{w(\mathbf{q})}{(2\pi)^2}\theta(|\mathbf{q} |-q_{\mathrm{min}})\delta(1-z) \right] \\
& \quad \times
\delta(x - yz)\, \delta(\mathbf{k} - \mathbf{q} -z\mathbf{k}') \,D(y,\mathbf{k}',\tau)\,,
\end{aligned}
\label{eq:ktee3}
\end{equation}
where we have introduced the upper (infra-red) cut-off for the $z$ integral 
and the lower cut-off $q_{\mathrm{min}}$ in the integral of 
$w(\mathbf{q})$ over $\mathbf{q}$, 
since the former is divergent for $z\rightarrow 1$ and the latter
is divergent for $|\mathbf{q}| \rightarrow 0$ in both cases of $w(\mathbf{q})$ given in Eqs.~(\ref{eq:wq1}) and  (\ref{eq:wq2}). 
The whole equation does not depend on these cut-offs, so they (if sufficiently small)
only play roles of (dummy) regulators of the corresponding integrals.

The expression for the integral $W$ depends on the actual form of the function $w(\mathbf{q})$.
For $w(\mathbf{q})$ given in Eq.~(\ref{eq:wq1}) we get
\begin{equation}
 W^{(1)} = \frac{4\pi\alpha_s^2 N_c n t^*}{q_{\mathrm{min}}^2}\,,
\label{eq:ktW1}
\end{equation}
while for the one of Eq.~(\ref{eq:wq2}):
\begin{equation}
 W^{(2)} = \frac{8\pi\alpha_s^2 N_c n t^*}{m_D^2}\,
 \ln\frac{\sqrt{q_{\mathrm{min}}^2+m_D^2}}{q_{\mathrm{min}}}\,.
\label{eq:ktW2}
\end{equation}
The latter features much milder dependence on the lower cut-off $q_{\mathrm{min}}$ than the former. 
The integral $\Phi(x)$ does not have a compact analytical form,
instead it can be computed numerically for a given value of $\epsilon$.
However, as will be seen later on, we do not need its actual value.

The above integro-differential equation can then be transformed 
into an integral equation
\begin{equation}
\begin{aligned}
 D(x,\mathbf{k},\tau) & =  e^{-\Psi(x)(\tau - \tau_0)} \, D(x,\mathbf{k},\tau_0) \\
& +  \int_{\tau_0}^{\tau} d\tau' \int_0^1 dz\, \int_0^1 dy\, \int d^2\mathbf{k'} \int d^2\mathbf{q} \; {\cal G}(z,\mathbf{q})\\
 &\quad \times \delta(x \; - \; zy) \, \delta(\mathbf{k} - \mathbf{q} - z\mathbf{k'})\, e^{-\Psi(x)(\tau - \tau')} \,D(y,\mathbf{k'},\tau')\,,
\end{aligned}
\label{eq:ktee4}
\end{equation}
where we have introduced the following notation
\begin{equation}
{\cal G}(z,\mathbf{q}) = \sqrt{\frac{z}{x}} \, z{\cal K}(z)\,\theta(1-\epsilon-z) \,\delta(\mathbf{q}) 
                                 +  t^* \, \frac{w(\mathbf{q})}{(2\pi)^2} \,\theta(|\mathbf{q} | - q_{\mathrm{min}}) \delta(1-z) \, .
\label{eq:ktKer}
\end{equation}
The factor $e^{-\Psi(x)(\tau - \tau')}$ is the Sudakov-type form factor 
corresponding to resummation of virtual and unresolved-real 
gluon emissions as well as unresolved collisions with the medium 
due to the kernel-function $C(\mathbf{q})$. 

Similarly, Eq.~(\ref{eq:eee1}) can be transformed into the integral equation
\begin{equation}
\begin{aligned}
D(x,\tau) =\; & e^{-\Phi(x)(\tau - \tau_0)} D(x,\tau_0) \\
               + & \int_{\tau_0}^{\tau} d\tau'  \int_0^{1-\epsilon}  dz \int_0^1 dy \, \delta(x - zy) \sqrt{\frac{z}{x}} \,z{\cal K}(z) e^{-\Phi(x)(\tau - \tau')} D(y,\tau') \,.
\end{aligned}
\label{eq:eee2}
\end{equation}
In this case the Sudakov-type form factor $e^{-\Phi(x)(\tau - \tau')}$ corresponds only 
to resummation of virtual and unresolved-real gluon emissions, as there is no medium-collision term.

The above integral equations can be formally solved by iteration.
For Eq.~(\ref{eq:ktee4}) we obtain
\begin{equation}
\begin{aligned}
 D(x,\mathbf{k},\tau)  =\:& \int_0^1 dx_0 \,\int d^2 \mathbf{k}_0 \, D(x_0,\mathbf{k}_0,\tau_0) \bigg\{ e^{-\Psi(x_0)(\tau - \tau_0)} \delta(x-x_0)\,\delta(\mathbf{k}-\mathbf{k}_0)\\
 +\:& \sum_{n=1}^{\infty}\prod_{i=1}^n \left[ \int_{\tau_{i-1}}^{\tau} d\tau_i \, \int_0^1 dz_i\, \int d^2\mathbf{q}_i\,
 {\cal G}(z_i,\mathbf{q}_i)\,e^{-\Psi(x_{i-1})(\tau_i - \tau_{i-1})} \right] \\
 &\times e^{-\Psi(x_n)(\tau - \tau_n)} \, \delta(x-x_n)\,\delta(\mathbf{k}-\mathbf{k}_n)\bigg\}\,,
\end{aligned}
\label{eq:ktItfin}
\end{equation}
where
\begin{equation}
x_n = z_n x_{n-1},\qquad 
\mathbf{k}_n = z_n \mathbf{k}_{n-1} + \mathbf{q}_n \,,
\label{eq:xnkn}
\end{equation}
with $x_0$ and $\mathbf{k}_0$ being some initial values of 
$x$ and $\mathbf{k}$ at the initial evolution time $\tau_0$, 
given by the distribution $D(x_0,\mathbf{k}_0,\tau_0)$.

A similar solution can be found for Eq.~(\ref{eq:eee2}):
\begin{equation}
\begin{aligned}
D(x,\tau) & =  \int_0^1 dx_0 \, D(x_0,\tau_0) \bigg\{  e^{-\Phi(x_0)(\tau - \tau_0)}\, \delta(x - x_0) \\
    & + \sum_{n=1}^{\infty} \prod_{i=1}^n  \left[ \int_{\tau_{i-1}}^{\tau} d\tau_i  \int_0^1  dz_i\, \sqrt{\frac{z_i}{x_i}} \,z_i{\cal K}(z_i)\, \theta(1-\epsilon-z_i)\,e^{-\Phi(x_{i-1})(\tau_i - \tau_{i-1})} \right]
    \\
    & \times e^{-\Phi(x_n)(\tau - \tau_n)} \, \delta(x - x_n) \bigg\}\,.
\end{aligned}
\label{eq:iterfin}
\end{equation}

\section{Markov Chain Monte Carlo algorithms}
\label{sec:MCMC}

The formal solutions given in Eqs.~(\ref{eq:ktItfin}) 
and (\ref{eq:iterfin}) can be used to develop Markov Chain Monte Carlo 
algorithms for numerical evaluation 
of the distribution functions $D(x,\tau)$ and $D(x,\mathbf{k},\tau)$, 
given some initial functions $D(x_0,\tau_0)$ and
$D(x_0,\mathbf{k}_0,\tau_0)$, respectively.

In Eqs.~(\ref{eq:ktItfin}) and (\ref{eq:iterfin}) there is 
ordering in the variable $\tau_i$:
\begin{equation}
\tau_0 < \tau_1 <  \tau_2 < \ldots < \tau_i < \ldots < \tau.
\label{eq:tauord}
\end{equation}
Therefore, this variable can be treated as an evolution time of a random walk in the MCMC algorithm
to be used for a numerical solution of the respective integral
equation.
This random walk will start at some time $\tau_0$ and
finish at the time moment $\tau$, making an arbitrary number
of random leaps between $\tau_{i-1}$ and 
$\tau_i,\; i = 1,2,\ldots$.
In order to construct such an algorithm let us rewrite this equation in a probabilistic form.

First we notice that
\begin{equation}
 e^{-\Psi(x_n)(\tau - \tau_n)} = \int_{\tau}^{+\infty} d\tau_{n+1}\,\Psi(x_n) e^{-\Psi(x_n)(\tau_{n+1} - \tau_n)},\quad n = 0,1,\ldots, 
\label{eq:stoping}
\end{equation}
and this can be regarded the probability of a single jump beyond $\tau$ from the point $\tau_n$, i.e.\ a stopping rule
for the random walk.
Thus, the probability density function ({\it pdf}) of a random variable $\tau_i$ for a single leap from $\tau_{i-1}$ is
\begin{equation}
 \varrho(\tau_i) = \Psi(x_{i-1}) e^{-\Psi(x_{i-1})(\tau_i - \tau_{i-1})},\quad \tau_i \in [\tau_{i-1},+\infty).
 \label{eq:probtau}
\end{equation}
The random variable $\tau_i$ can be generated according to
the above {\it pdf} using the analytical inverse transform method.

The {\it pdf} for the variables $z_i$ and $\mathbf{q}_i$ is given by
\begin{equation}
 \xi(z_i,\mathbf{q}_i) = \frac{{\cal G}(z_i,\mathbf{q_i})}{\Psi(x_{i-1})}\, .
 \label{eq:probzq}
\end{equation}
The variables $z_i$ and $\mathbf{q}_i$ can be generated almost independently using the following branching MC method:
\begin{itemize}

\item 
with the probability $p = \Phi(x_{i-1})/\Psi(x_{i-1})$ generate $z_i$ according to the density function $\zeta(z_i)$:
\begin{equation}
 \zeta(z_i) = \frac{z_i{\cal K}(z_i)}{\kappa(\epsilon)}\, ,
 \qquad 
 \kappa(\epsilon) = \int_0^{1-\epsilon} dz\,z{\cal K}(z),
 \label{eq:probz}
\end{equation}
and set $\mathbf{q}_i = \mathbf{0}$, 
\item
otherwise, i.e.\ with the probability $1-p$, set $z_i=1$ 
and generate $\mathbf{q}_i$ according to the density function
\begin{equation}
 \omega(\mathbf{q}_i) = \frac{1}{W}\,t^* \, \frac{w(\mathbf{q}_i)}{ (2\pi)^2} \,, \quad |\mathbf{q}_i| \ge q_{\mathrm{min}}\, .
\label{eq:probq}
\end{equation}

\end{itemize}
 While the random variable $\mathbf{q}_i$ can be generated
according to the {\it pdf} $\omega(\mathbf{q}_i)$ for
$w(\mathbf{q})$ given in Eqs.~(\ref{eq:wq1}) and 
(\ref{eq:wq2}), generating $z_i$ according to
the {\it pdf} $\zeta(z_i)$ is more difficult due to 
the complicated function ${\cal K}(z)$ given in
Eq.~(\ref{eq:kernel1}). 
For this purpose one can use the rejection method
or the importance sampling -- we shall come back to this
later on.

Finally, let us define the probability density for the initial variables $x_0$ and $\mathbf{k}_0$:
\begin{equation}
 \eta(x_0,\mathbf{k}_0) = \frac{D(x_0,\mathbf{k}_0,\tau_0)}{d(\tau_0)} , \quad d(\tau_0) = \int_0^1 dx_0\, \int d^2\mathbf{k}_0\,D(x_0,\mathbf{k}_0,\tau_0).
 \label{eq:probx0kt0}
\end{equation}
If this function is complicated, for generation of the random variables $x_0$ and $\mathbf{k}_0$ one can use
some self-adaptive Monte Carlo (MC) sampler, e.g.\ 
\foam~\cite{Jadach:2002kn}.
However, quite often it factorises into a product of probability densities:
 \begin{equation}
 \eta(x_0,\mathbf{k}_0) = \chi(x_0)\upsilon(\mathbf{k}_0)\,,
 \label{eq:factprobx0kt0}
\end{equation}
where for $\upsilon(\mathbf{k}_0)$ one can use e.g.\ the Gaussian distribution:
 \begin{equation}
 \upsilon(\mathbf{k}_0) = \frac{1}{2\pi\sigma_{\mathbf{k}_0}^2}\, \exp\left[-\frac{\mathbf{k}_0^2}{2\sigma_{\mathbf{k}_0}^2}\right] \,,
 \label{eq:probkt0}
\end{equation}
which can be easily generated, e.g.\ using the Box--Muller 
method.

Having defined all the necessary probability distribution functions, we can rewrite Eq.~(\ref{eq:ktItfin}) in the following form
\begin{equation}
\begin{aligned}
D(x,\mathbf{k},\tau) = &\: d(\tau_0) \int_0^1 dx_0 \int d^2\mathbf{k}_0\,  \eta(x_0,\mathbf{k}_0)  \bigg\{   \int_{\tau}^{+\infty} d\tau_1\,\varrho(\tau_1) \, 
\delta(x - x_0)\,\delta(\mathbf{k}-\mathbf{k}_0)\\
    & + \sum_{n=1}^{\infty} \prod_{i=1}^n  \left[ \int_{\tau_{i-1}}^{\tau} d\tau_i \, \varrho(\tau_i) \int_0^1  dz_i  \int d^2\mathbf{q}_i \; \xi(z_i,\mathbf{q}_i) \right] 
      \int_{\tau}^{\infty} d\tau_{n+1}\,\varrho(\tau_{n+1}) \\
   & \times   \delta(x - x_n)\, \delta(\mathbf{k}-\mathbf{k}_n) \bigg\}  .
\end{aligned}
\label{eq:ktiterprob}
\end{equation}

Now we can propose the following MCMC algorithm for numerical numerical evaluation of  Eq.~(\ref{eq:ktiterprob}):
\begin{description}
\item[Step 1]
Start a random walk from the point $\tau_0$. First generate the variables $x_0\in [0,1]$ and $\mathbf{k}_0$ according to the probability
density $\eta(x_0,\mathbf{k}_0)$, then generate $\tau_1 \in [\tau_0,+\infty)$ according to the probability density $\varrho(\tau_1)$.
If $\tau_1 > \tau$, set $x=x_0,\,\mathbf{k}=\mathbf{k}_0$ and {\bf stop} the random walk, otherwise go to {\bf step~2}.
\item[Step 2]
Generate the variables $z_1\in[0,1]$ and $\mathbf{q}_1> \mathbf{q}_{\mathrm{min}}$ according to the probability density $\xi(z_1,\mathbf{q}_1)$ and calculate 
$x_1=z_1x_0,\, \mathbf{k}_1=\mathbf{q}_1 +z_1\mathbf{k}_0$. 
Then generate $\tau_2 \in [\tau_1,+\infty)$ according to the probability density $\varrho(\tau_2)$:
if $\tau_2 > \tau$, set $x=x_1,\,\mathbf{k}=\mathbf{k}_1$ and {\bf stop} the random walk, otherwise go to {\bf step~3}.
\item[...]
\item[Step n]
Generate the variables $z_n\in[0,1]$ and $\mathbf{q}_n> \mathbf{q}_{\mathrm{min}}$ according to the probability density $\xi(z_n,\mathbf{q}_n)$  
and calculate $x_n=z_nx_{n-1},\,\mathbf{k}_n=\mathbf{q}_n +z_n\mathbf{k}_{n-1}$. 
Then generate $\tau_{n+1} \in [\tau_n,+\infty)$ according to the probability density $\varrho(\tau_{n+1})$:
if $\tau_{n+1} > \tau$, set $x=x_n,\mathbf{k}=\mathbf{k}_n$ and {\bf stop} the random walk, otherwise go to {\bf step $\bf n+1$}.
\item[...]
\end{description}
{\bf Repeat} the above steps $N$-times histogramming the variables $x$ and $\mathbf{k}$. 
At the end normalise the histograms with the value $d(\tau_0)/N$. 
Such a 3D distribution of $x$ and  $\mathbf{k}$
will be a Monte Carlo estimate of the function $D(x,\mathbf{k},\tau)$ for a given value of $\tau$
with a statistical error proportional to $1/\sqrt{N}$.
Since a 3D distribution is difficult to visualise, in practice one usually makes 1D or 2D 
histograms of any combination of $x$  and $\mathbf{k}$. 
In addition, one can impose arbitrary cuts on any of these variables. 

One can formally prove that the above algorithm gives a correct
solution to Eq.~(\ref{eq:ktiterprob}), 
i.e.\ that the expectation value of a MC weight 
associated with a random walk trajectory, as described above, 
is equal to the function $D(x,\mathbf{k},\tau)$. 
We skip such a proof here -- it will be provided in our future publication dedicated to the MCMC algorithm
and its implementation.

In the above MCMC algorithm we have assumed that all random variables 
can be generated according to the respective probability 
distribution functions using standard Monte Carlo techniques, 
preferably the analytical inverse-transform method or its combination with the branching method.  
Among the integration variables in Eq.~(\ref{eq:ktiterprob}) the most problematic is the variable $z$ because 
its probability distribution function $\zeta(z)$ is too 
complicated to be sampled with the above methods.
Details on how to deal with this using a combination of the
branching method with the importance sampling are given in Appendix~A.

The MCMC algorithm for solving Eq.~(\ref{eq:iterfin})
is analogous to the above 
-- one only needs to set $w(\mathbf{q}) = 0$ and 
$\mathbf{k}_n = \mathbf{0},\, n = 0,1,\ldots$.

\section{Differential method}
\label{sec:DifMet}

Direct temporal numerical integration of the Eq.~(\ref{eq:eee1}) is another approach we use. First of all, the spatial
grid with the constant step-size $\Delta x = 1/N $ is created to keep $N$ grid-points with the solution of $D(x,\tau)$ 
at each time-step $\Delta \tau$. The generation of this grid seems to be quite involved in order to obtain reasonable
numerical results and we shall discuss this issue later on. The integral on the right-hand side of the
Eq.~(\ref{eq:eee1}) is divided into two parts: the gain part
\begin{equation}
\int_x^1 dz\, {\cal K}(z) \sqrt{\frac{z}{x}}\, D\left(\frac{x}{z},\tau\right) 
\label{eq:Ngi}
\end{equation}
and loss part
\begin{equation}
-\int_0^1 dz\, {\cal K}(z) \frac{z}{\sqrt{x}}\, D(x,\tau). 
\end{equation}
Both of them are evaluated at every time step by a simple midpoint rule, i.e.
\begin{equation}
\int_{z-\frac{\Delta z}{2}}^{z+\frac{\Delta z}{2}} dz\, f(z,x,\tau) = \Delta z f(z,x,\tau),
\end{equation}
where $\Delta z$ is the spatial step-size equal to $\Delta x$. Other advanced methods\footnote{Simpson's $3/8$ rule,
advanced integrators from the {\tt QUADPACK} package and also Monte Carlo integrators available in the GNU Scientific Library~\cite{Galassi:2009} were tested on moderate ($N$ up to 1000 grid-points) grids.} did not yield significantly better results and this type of interpolation function is the fastest choice, which in turn allows us to use very dense
grids (in fact, the grids with as many as $N=16384$ grid-points were used to obtain numerical results that are reasonably
accurate and are presented in this paper). Due to the simple midpoint approximation we were able to keep computational time less than one day on a computer system with the i7 CPU.

After the spatial approximation, we end up with the following system of ordinary differential equations:
\begin{equation}
\frac{d}{d\tau}D(x_i,\tau)=\sum_{j\geq i}^{N-1}{\cal K}(z_j)\sqrt{\frac{z_j}{x_i}}D\left(\frac{x_i}{z_j},\tau\right)\Delta x-\sum_{j=0}^{N-1}
{\cal K}(z_j)\frac{z_j}{\sqrt(x_i)}D(x_i,\tau),
\label{eq:Nae}
\end{equation}
with $x_i=(i+0.5)/N$ and $z_j=(j+0.5)/N$, $i,j\in\{0,\ldots,N-1\}$.

The RHS in Eq.~(\ref{eq:Nae}) is then used to advance the solution in time by the five-stage Runge--Kutta--Merson method 
with fourth-order accuracy and adaptive time-step regulator, see e.g.~\cite{Christiansen:1970}. 
The time step correction is accomplished by the following rule:
\begin{equation}
\Delta t_\mathrm{new} = 0.8\Delta t_\mathrm{old}\left( \frac{\delta_\mathrm{tol}}{\varepsilon}\right)^{\frac{1}{5}},
\end{equation}
where $\delta_\mathrm{tol}=10^{-12}$ is the tolerance parameter used in the simulations 
and $\varepsilon$ is the truncation error indicator computed in the last step of the algorithm.

The initial condition used in the simulations is the analytical solution (\ref{eq:Smplf}) at the very small time
$\tau_0=10^{-4}$ which we use as an approximation of the $\delta$-function in the case of the simplified kernel. In the full kernel case, we use the following approximation of the $\delta$-function:
\begin{equation}
    D(x,0)=\frac{1}{\varepsilon}\exp\left[-\left(\frac{1-x}{\varepsilon}\right)^2\right],
\end{equation}
with $\varepsilon=6\cdot10^{-3}$.

The numerical solution is then advanced 
in time and reported. Spatial grids used during the simulations have to be very fine to obtain stable and meaningful
results. The problem lies in the gain integral (\ref{eq:Ngi}) where the arguments of ${\cal K}$ and $D$ are reciprocal.
When we use the equally distributed fixed grid and try to compute the gain integral with $z$ from the finite subset of
grid-points, we get arguments for $D$ from unequally distributed grid points due to $x/z$, 
i.e.\ a bad approximation of the gain integral as most of the values will be taken from the region close to $0$.
A substitution does not help here as it will just switch the reciprocal values from one term to the other. 
To overcome this problematic behaviour, the very fine grid is needed that has an inevitable effect on computational times.
One possible solution is the adaptive mesh refinement together with a smart distribution of the grid-points 
-- this type of approach is still investigated.

\section{Numerical results}
\label{sec:NumRes}

\begin{figure}[!ht]
\setlength{\unitlength}{0.1mm}
  \begin{picture}(1600,1200)
   \put(   0,600){\makebox(0,0)[lb]{\includegraphics[width=80mm, height=60mm]{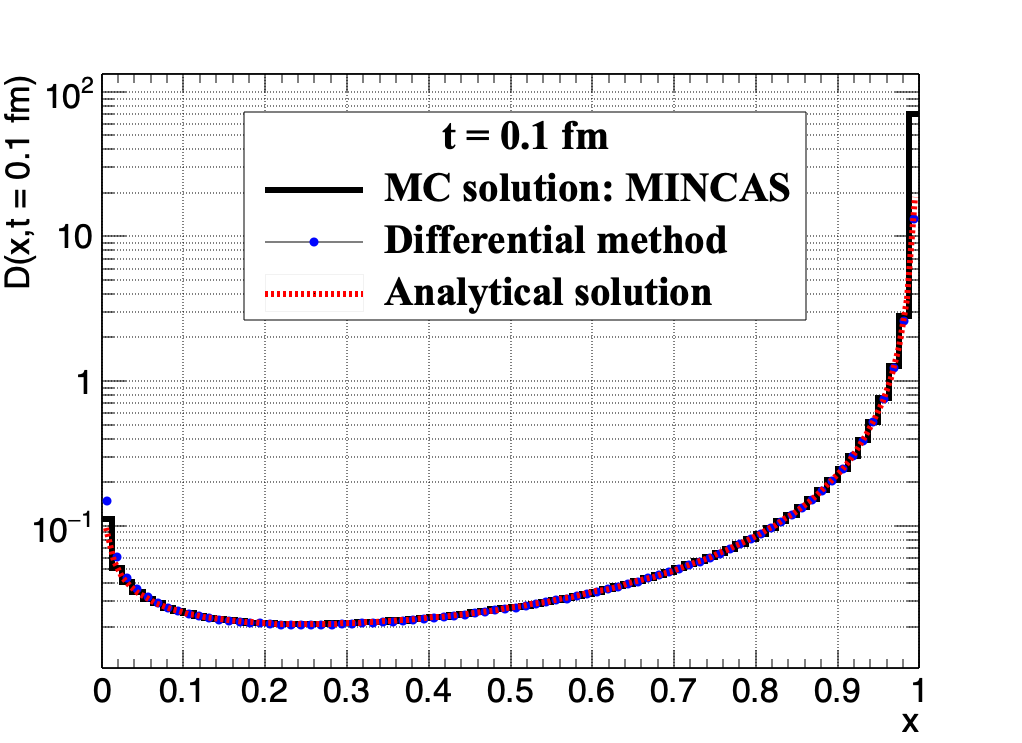} }} 
   \put(800,600){\makebox(0,0)[lb]{\includegraphics[width=80mm, height=60mm]{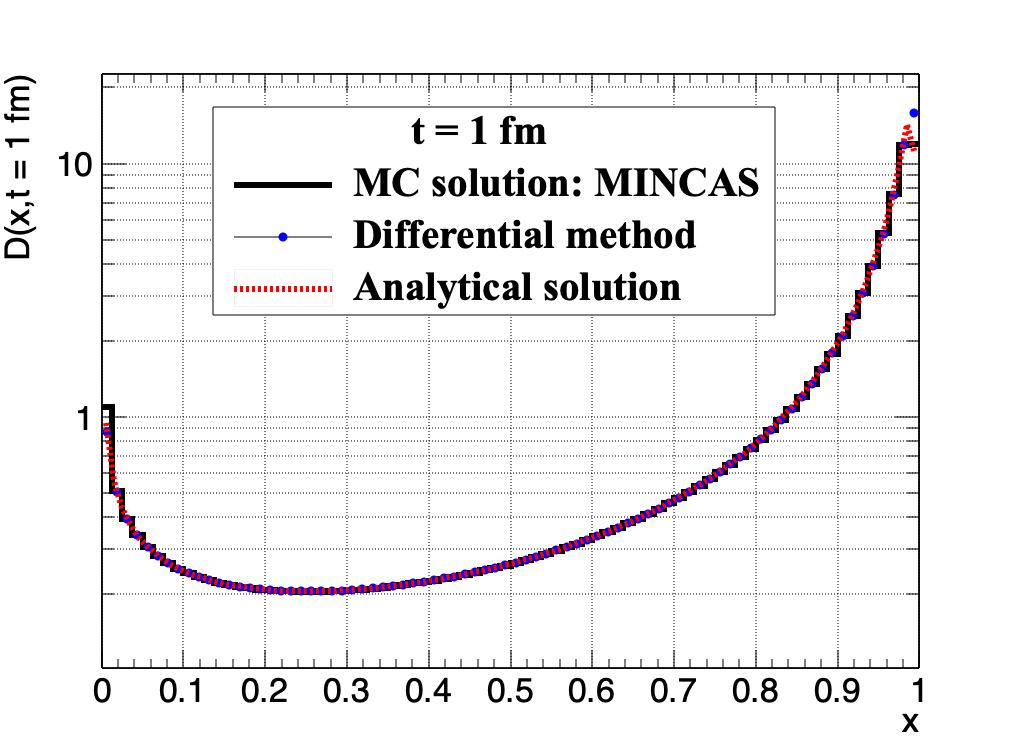} }}
   \put(   0,     0){\makebox(0,0)[lb]{\includegraphics[width=80mm, height=60mm]{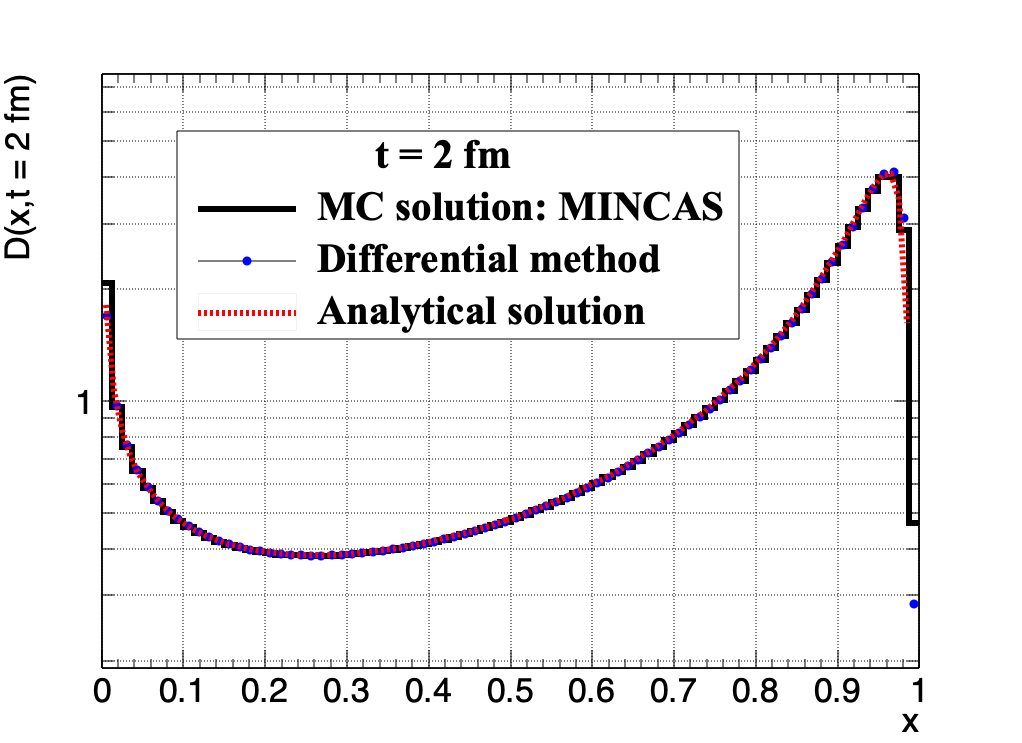} }}
   \put(800,    0){\makebox(0,0)[lb]{\includegraphics[width=80mm, height=60mm]{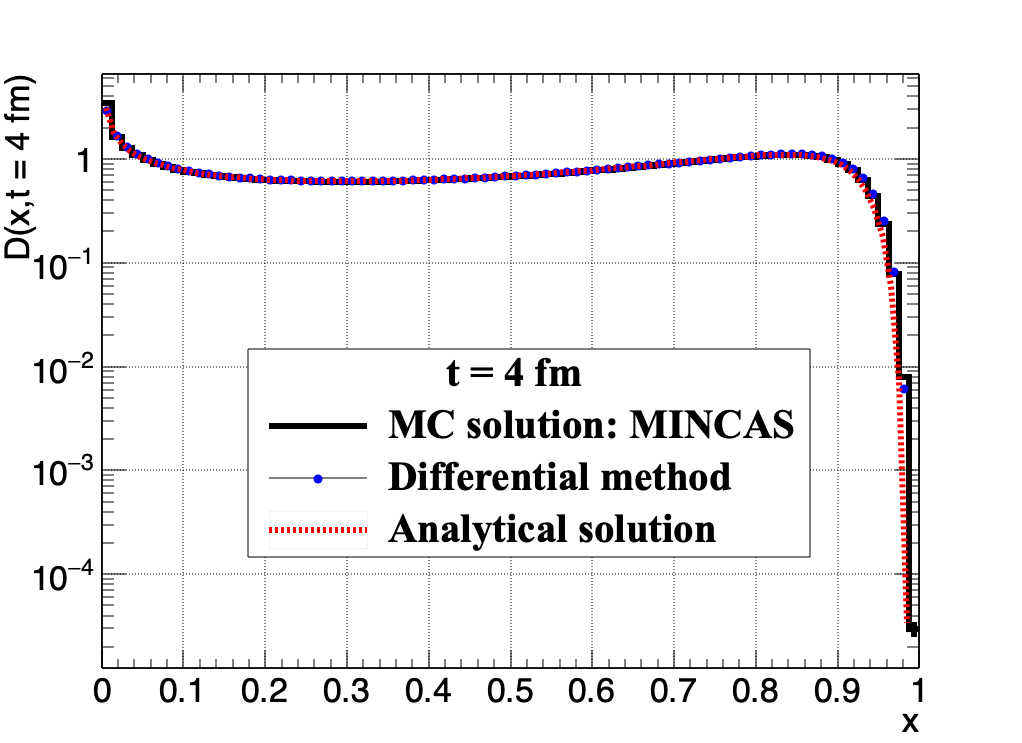} }}
 \end{picture}
\caption{\sf 
  Comparisons of the $x$ distributions from the Monte Carlo program 
  \mincas\ for the solution of Eq.~(\ref{eq:iterfin}) in the case of 
  the simplified $z$-kernel function with the
  differential method results and the analytical formula of 
  Eq.~(\ref{eq:Smplf}), for the evolution time values: 
  $t = 0.1,\,1,\,2\,,4\,$fm.
}
\label{fig:xdis_skz}
\end{figure}

We have implemented the MCMC algorithms described 
in Section~\ref{sec:MCMC} in the {\sf C}-language program 
called \mincas\ (the acronym for Medium-INduced CAScades) as two independent MC generators.
First, we performed numerical tests of the algorithm for the solution of 
Eq.~(\ref{eq:iterfin}) by comparing it with the analytical formula of Eq.~(\ref{eq:Smplf})
and with the numerical differential method described in Section~\ref{sec:DifMet} 
for the case of the simplified $z$-kernel function, i.e.\ with $f(z) = 1$.

In our numerical calculations presented below we have used the
following input parameters values:
\begin{eqnarray}
x_{\rm min} &=& 10^{-4}, \qquad \epsilon = 10^{-4} ,
\label{eq:xpar2} \\
q_{\rm min} &=& 0.1 \,{\rm GeV}, \qquad m_D = 0.993 \,{\rm GeV}, \qquad \sigma_{\mathbf{k}_0} = 0.1 \,{\rm GeV},
\label{eq:qkpar2} \\
N_c &=& 3, \quad \bar{\alpha} = 0.3,
\label{eq:albar} \\
E &=& 100\,{\rm GeV}, \quad n = 0.243 \,{\rm GeV}^{3}, \qquad \hat{q} = 1 \,{\rm GeV^2/fm}\, .
\label{eq:nqhpar}
\end{eqnarray}

The results for the evolution time values: $t = 0.1$, $1$, $2$ and 
$4\,$fm are presented in Fig.~\ref{fig:xdis_skz}. 
We can see a very good agreement between the three solutions: 
by the MCMC algorithm of \mincas, 
by the analytical formula (\ref{eq:Smplf}) obtained in \cite{Blaizot:2013hx} 
and by the differential method described in Section~\ref{sec:DifMet}.
The resulting distributions feature the turbulent behaviour, 
i.e.\ the energy is transported from the large-$x$ region to the low-$x$ region
without accumulating in the intermediate values of $x$.

\begin{figure}[!ht]
\setlength{\unitlength}{0.1mm}
  \begin{picture}(1600,1200)
   \put(   0,600){\makebox(0,0)[lb]{\includegraphics[width=80mm, height=60mm]{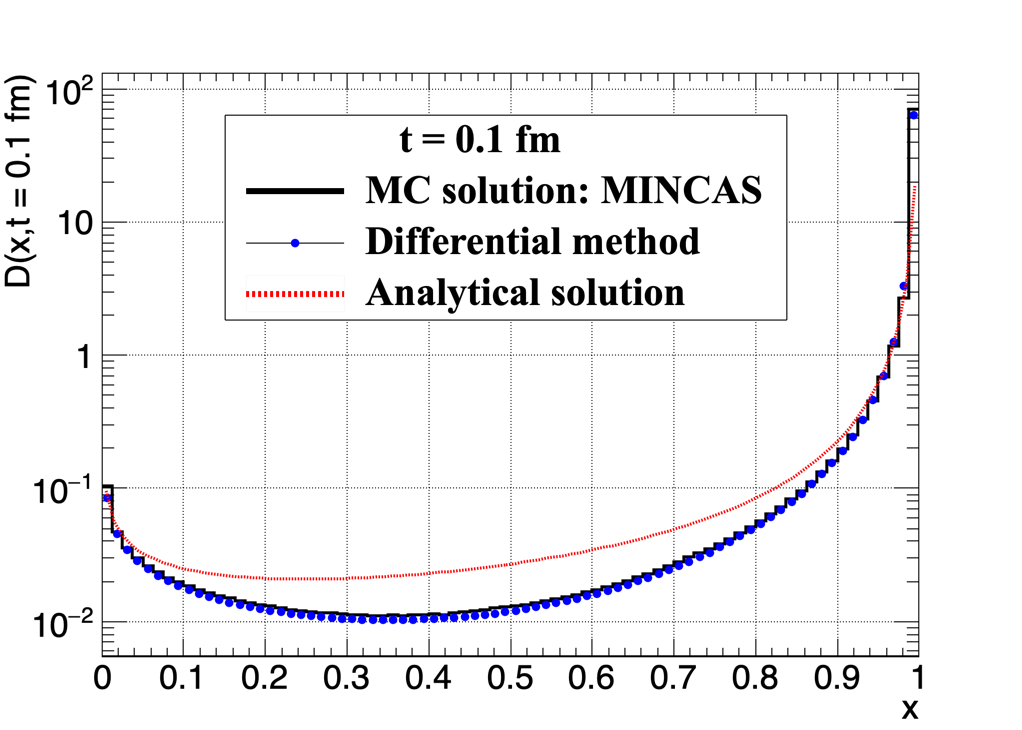} }} 
   \put(800,600){\makebox(0,0)[lb]{\includegraphics[width=80mm, height=60mm]{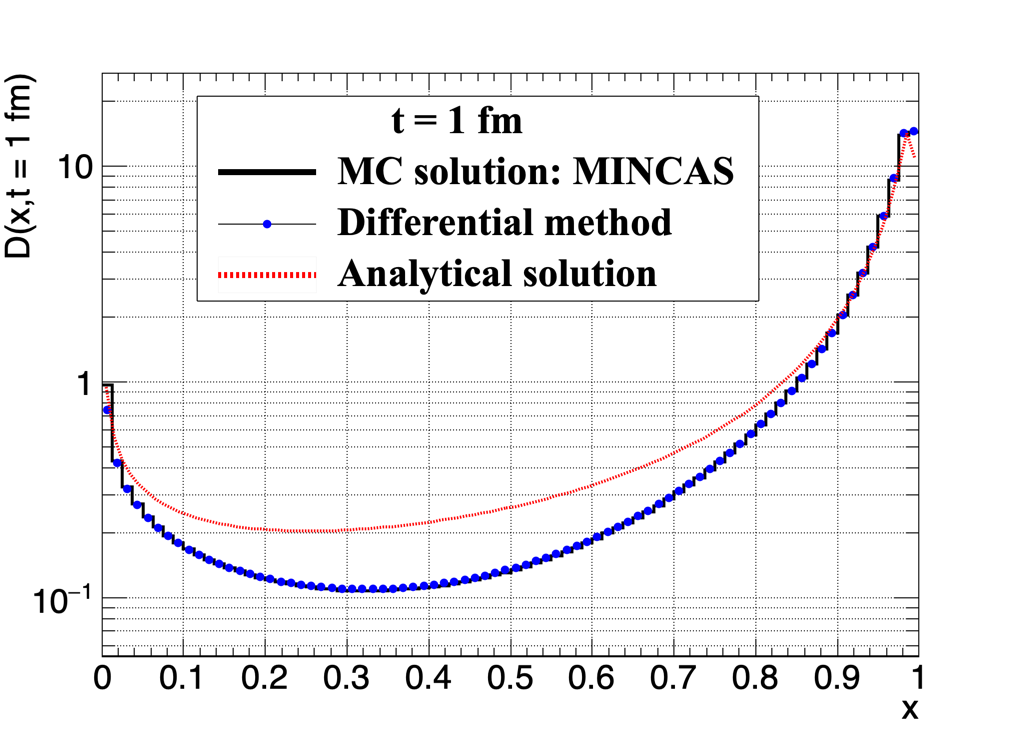} }}
   \put(   0,     0){\makebox(0,0)[lb]{\includegraphics[width=80mm, height=60mm]{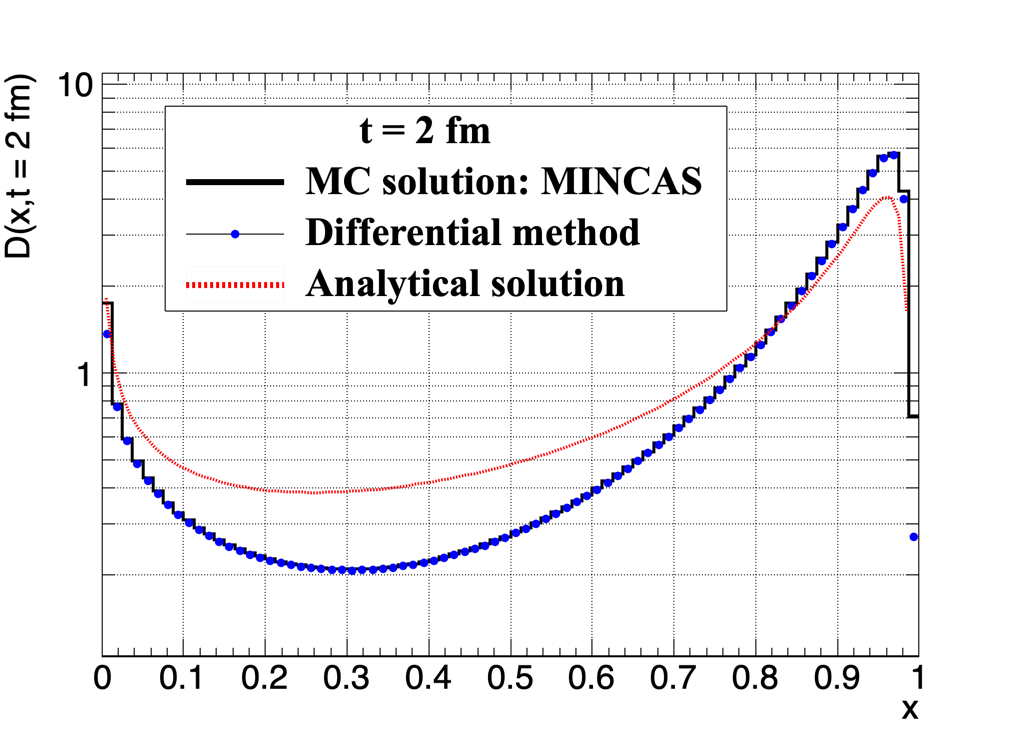} }}
   \put(800,    0){\makebox(0,0)[lb]{\includegraphics[width=80mm, height=60mm]{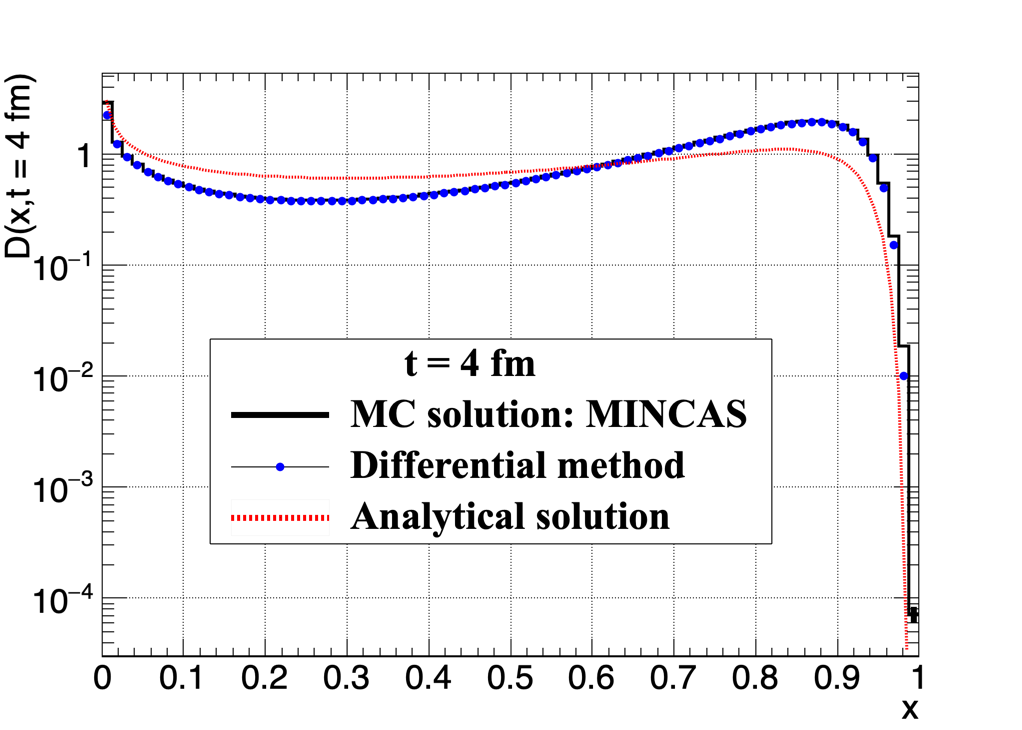} }}
 \end{picture}
\caption{\sf 
  Comparisons of the $x$ distributions from the Monte Carlo program 
  \mincas\ for the solution of Eq.~(\ref{eq:iterfin}) in the case of 
  the exact $z$-kernel function with the
  differential method results and the analytical formula of 
  Eq.~(\ref{eq:Smplf}), for the evolution time values: 
  $t = 0.1,\,1,\,2\,,4\,$fm.
}
\label{fig:xdis_fkz}
\end{figure}

In Fig.~\ref{fig:xdis_fkz} we show similar results as above, but
for the exact $z$-kernel function as given in Eq.~(\ref{eq:kernel1}).
The agreement between \mincas\ and the differential method is
similar as in Fig.~\ref{fig:xdis_skz} which confirms that
our numerical solutions of Eq.~(\ref{eq:eee1}) are also correct 
for the exact $z$-kernel. Of course, now the
analytical solution is away from both of them because it works only for
the simplified $z$-kernel -- it is shown only for reference.
One can see that the $x$ distribution for the exact $z$-kernel
differs considerably from the one for the simplified $z$-kernel,
particularly in the region of the intermediate $x$ values
-- the turbulent behaviour of the exact solution is stronger than
of the approximate one.


Then, we performed tests of the MCMC algorithm for the for the $x$ and $\mathbf{k}$
evolution of Eq.~(\ref{eq:ktItfin}) implemented in \mincas. 

\begin{figure}[t!]
\setlength{\unitlength}{0.1mm}
  \begin{picture}(1600,1200)
   \put(   0,600){\makebox(0,0)[lb]{\includegraphics[width=80mm, height=60mm]{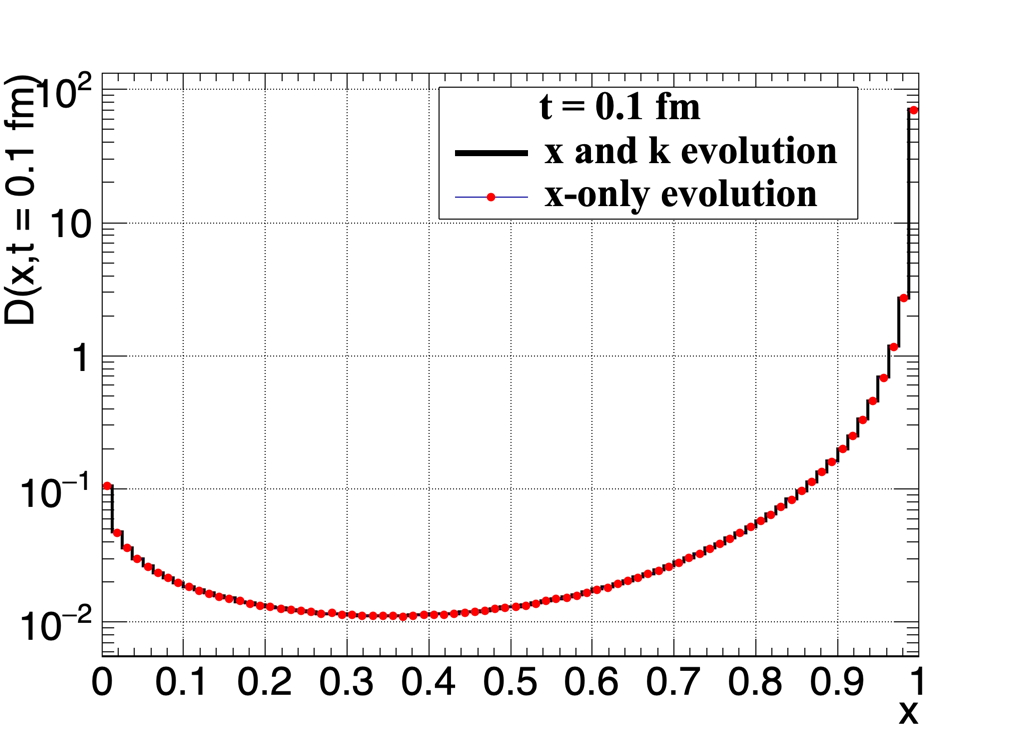} }} 
   \put(800,600){\makebox(0,0)[lb]{\includegraphics[width=80mm, height=60mm]{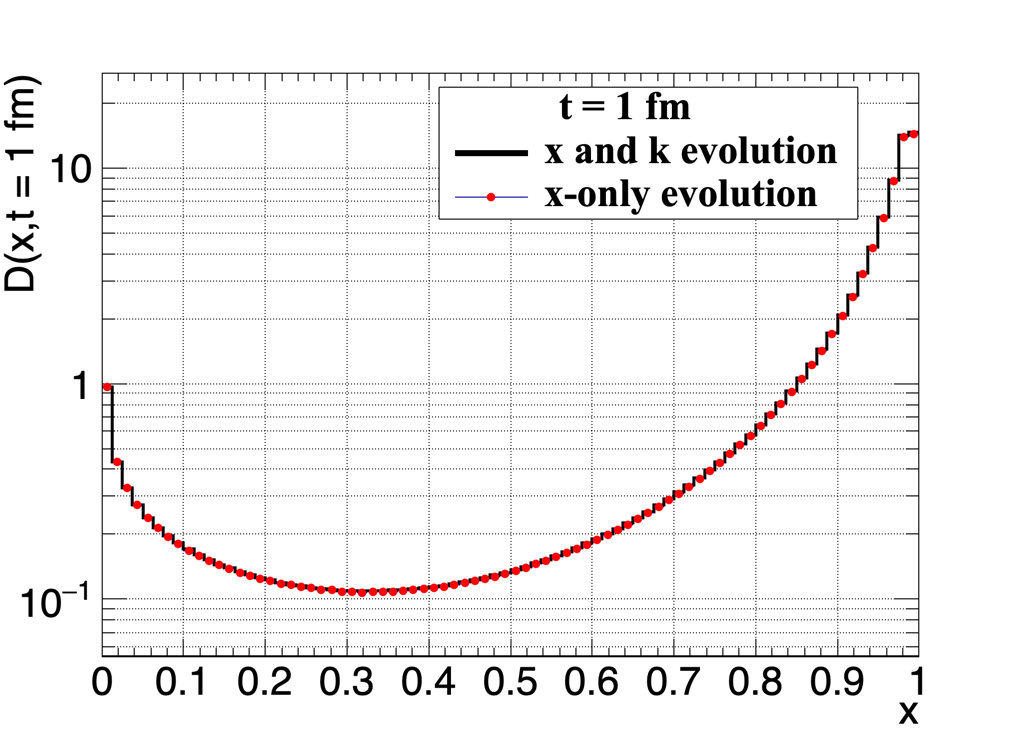} }}
   \put(   0,     0){\makebox(0,0)[lb]{\includegraphics[width=80mm, height=60mm]{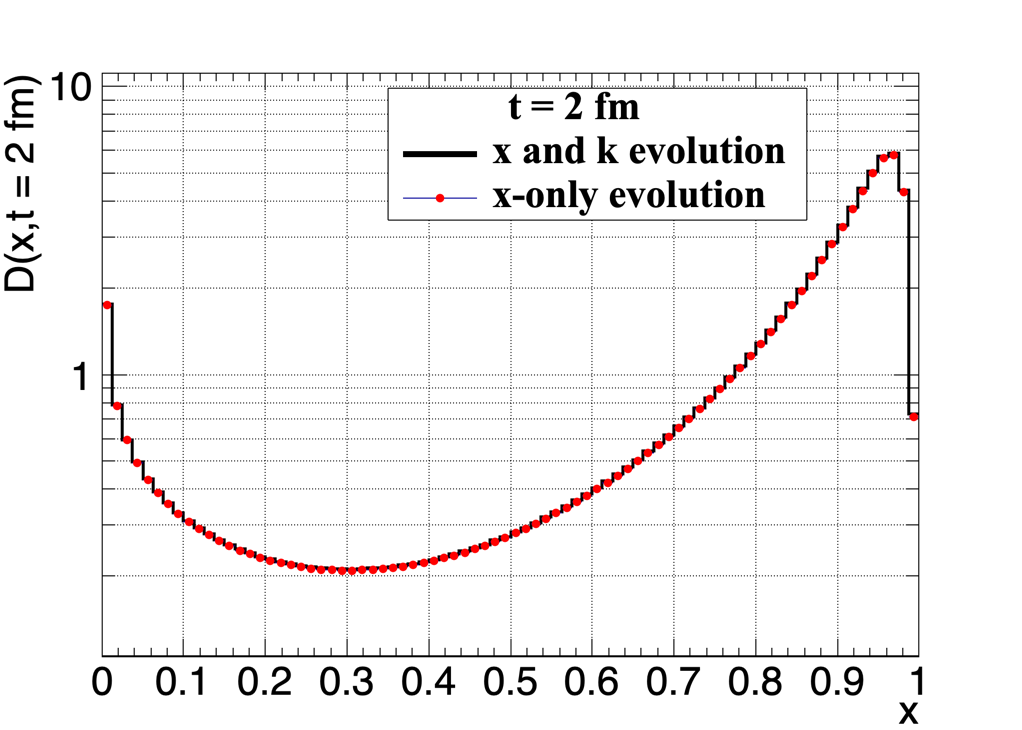} }}
   \put(800,    0){\makebox(0,0)[lb]{\includegraphics[width=80mm, height=60mm]{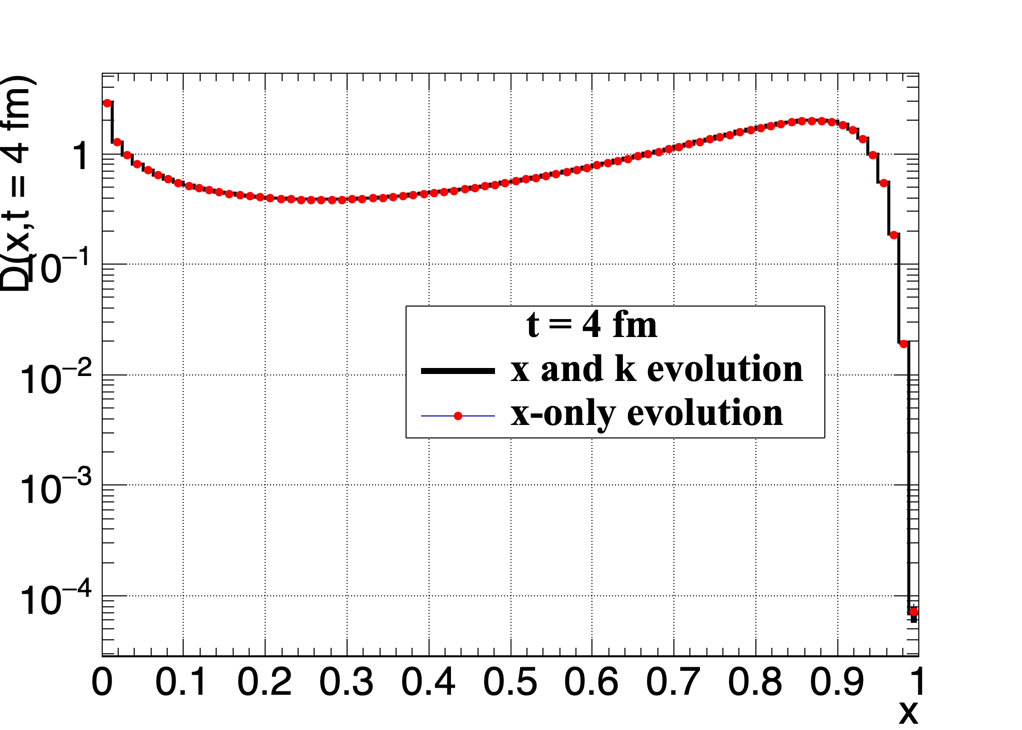} }}
 \end{picture}
\caption{\sf 
  Comparisons of the inclusive $x$ distributions from the Monte Carlo 
  program \mincas\ for the $x$ and $\mathbf{k}$ evolution 
  of Eq.~(\ref{eq:ktItfin})
  with the $x$-only evolution of Eq.~(\ref{eq:iterfin}), 
  for the evolution time values: $t = 0.1$, $1$, $2$ and $4\,$fm.
}
\label{fig:xdis_xk}
\end{figure}

Since the integration over $\mathbf{k}$ of Eq.~(\ref{eq:ktee1}) gives Eq.~(\ref{eq:eee1}), 
our first test was to check if using 
the algorithm for the $x$ and $\mathbf{k}$ evolution 
of Eq.~(\ref{eq:ktItfin})
we can reproduce the $x$ distributions generated by the
simpler algorithm for $x$-only evolution of Eq.~(\ref{eq:iterfin}).
For this purpose we have produced inclusive histograms of $x$, 
i.e.\ without any restrictions on $\mathbf{k}$. The results
in the case of the exact $z$-kernel function for the evolution time values: $t = 0.1$, $1$, $2$ and $4\,$fm 
are shown in Fig.~\ref{fig:xdis_xk}.
As one can see, they are in a perfect agreement. 
This is an important, non-trivial test of the MCMC algorithm for the $x$ and $\mathbf{k}$ evolution
of Eq.~(\ref{eq:ktItfin}) and its implementation in \mincas,
showing that it produces the correct $x$ distribution.

\begin{figure}[t!]
\setlength{\unitlength}{0.1mm}
  \begin{picture}(1600,1200)
   \put(   0,600){\makebox(0,0)[lb]{\includegraphics[width=80mm, height=60mm]{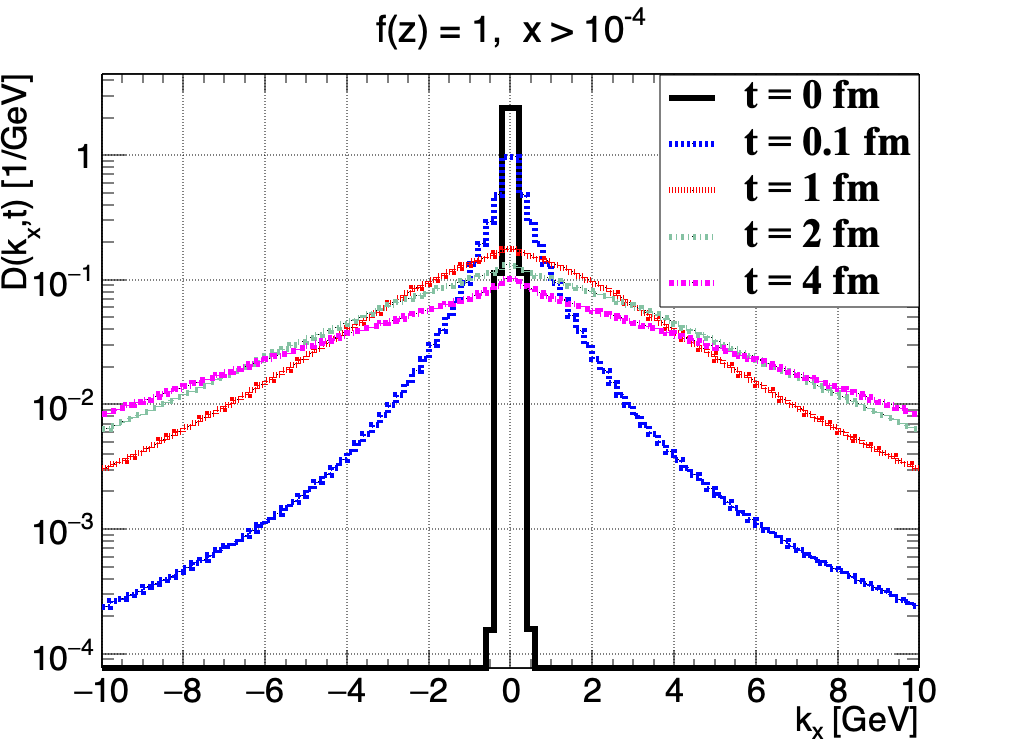} }} 
   \put(800,600){\makebox(0,0)[lb]{\includegraphics[width=80mm, height=60mm]{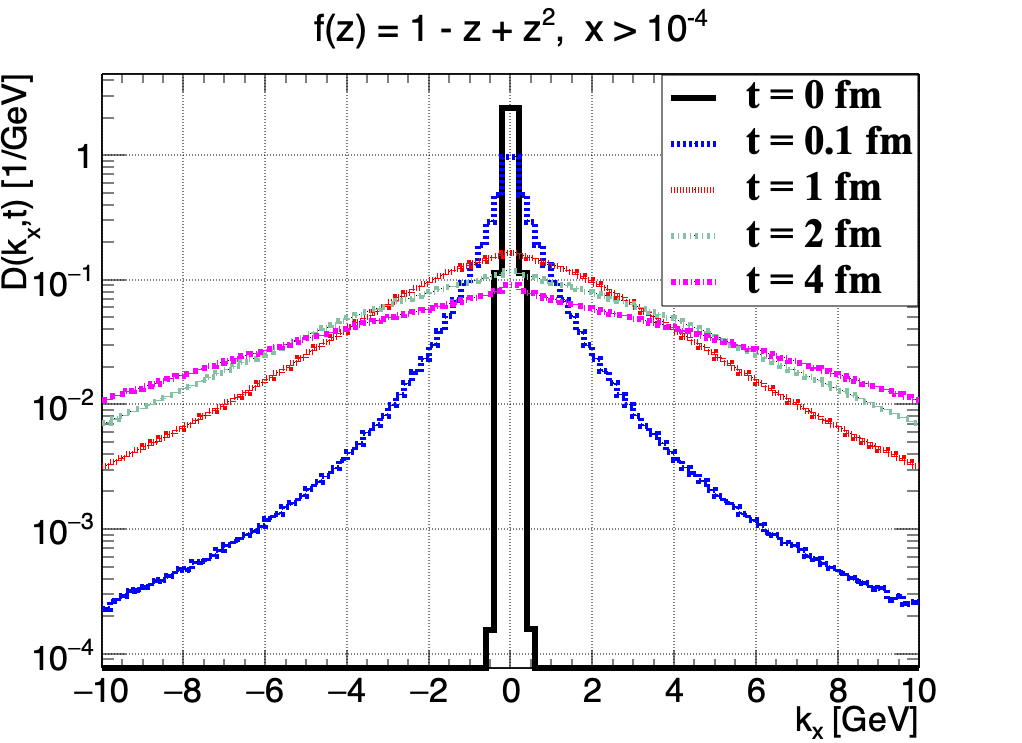} }}
   \put(   0,     0){\makebox(0,0)[lb]{\includegraphics[width=80mm, height=60mm]{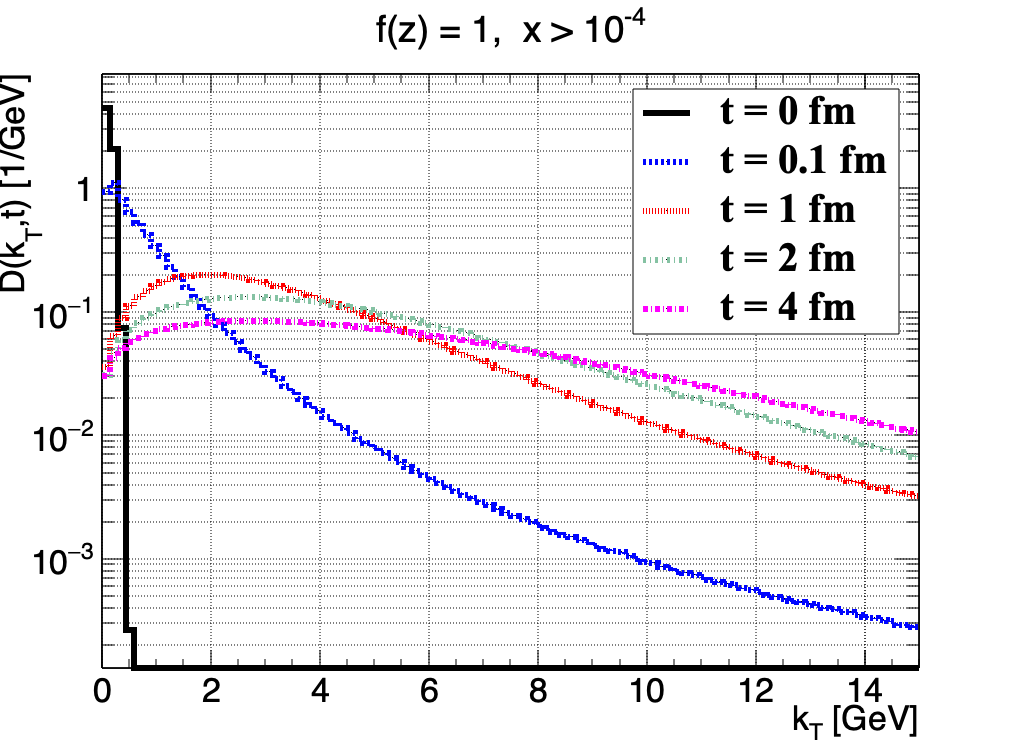} }}
   \put(800,    0){\makebox(0,0)[lb]{\includegraphics[width=80mm, height=60mm]{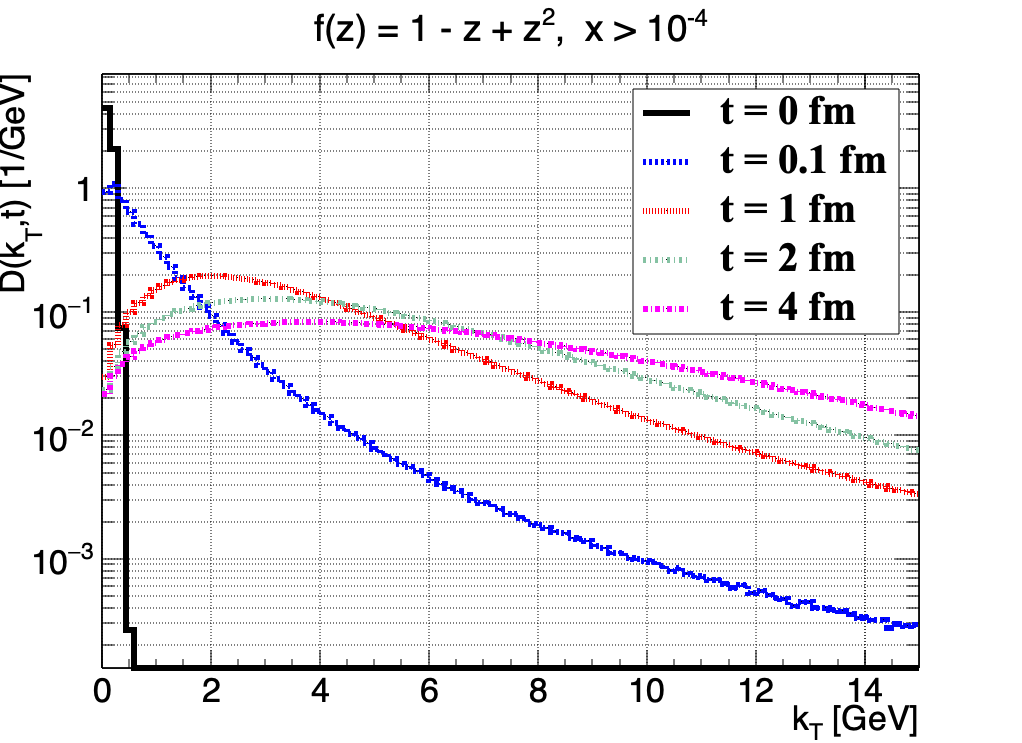} }}
 \end{picture}
\caption{\sf 
   The $k_x$ and $k_T$ distributions from the Monte Carlo program \mincas\ for the evolution time values: $t = 0,\, 0.1,\,1,\,2\,,4\,$fm.
   The LHS figures are for the simplified $z$-kernel function, while the RHS ones for the exact one.
}
\label{fig:kxkTdis}

\end{figure}
 
Unfortunately, we could not make comparisons of the $\mathbf{k}$ 
distributions with the differential method  because it turned out 
to be inefficient in solving the general evolution equation 
(\ref{eq:ktee1}). Therefore, in the following we present a few figures
with the results from \mincas\ only, to show how
the the medium-induced QCD evolution affects transverse gluon momenta.

In Fig.~\ref{fig:kxkTdis} the $k_x$ and $k_T=\sqrt{k_x^2+k_y^2}$ distributions are shown.
These results have been obtained with $w(\mathbf{q})$ of Eq.~(\ref{eq:wq2}) for the evolution time values: 
$t = 0$, $0.1$, $1$, $2$ and $4\,$fm. 
One can observe fast broadening of a very narrow initial Gaussian distribution of $\mathbf{k}$ with the increasing evolution time as well as the departure from the Gaussianity of the subsequent distributions. 
This non-Gaussian behaviour of the transverse momentum distributions can
be explained by inspecting Eq.~(\ref{eq:xnkn}). As one can see,
$\mathbf{k}_n$ is
the sum of $n+1$ random variables. From the Central Limit Theorem it
follows that for a fixed value of $n$ a distribution of the random 
variable $\mathbf{k}_n$ would converge to the Gaussian distribution. 
However, in this case $n$ is also
a random variable as it corresponds to the length of the random-walk
trajectory in the MCMC algorithm described in Section~\ref{sec:MCMC}.
The final distribution of $\mathbf{k}$ results from summing of
all such trajectories, therefore it is not a single Gaussian distribution
but a sum of an arbitrary number of Gaussian distributions with the same
mean values and different widths (variances). Generally, the longer trajectory results
in the larger width as the transverse momentum broadening due the medium-collisions
seems to dominate, for a given form of the function $w(\mathbf{q})$, 
over its shrinking due to the emission branchings.
As the evolution time increases the trajectories get longer
and more Gaussian distributions with larger widths contribute to
the overall $\mathbf{k}$ distribution, making it wider  
-- this we observe in Fig.~\ref{fig:kxkTdis}.

\begin{figure}[t!]
\setlength{\unitlength}{0.1mm}
  \begin{picture}(1600,600)
   \put(  0,0){\makebox(0,0)[lb]{\includegraphics[width=80mm, height=60mm]{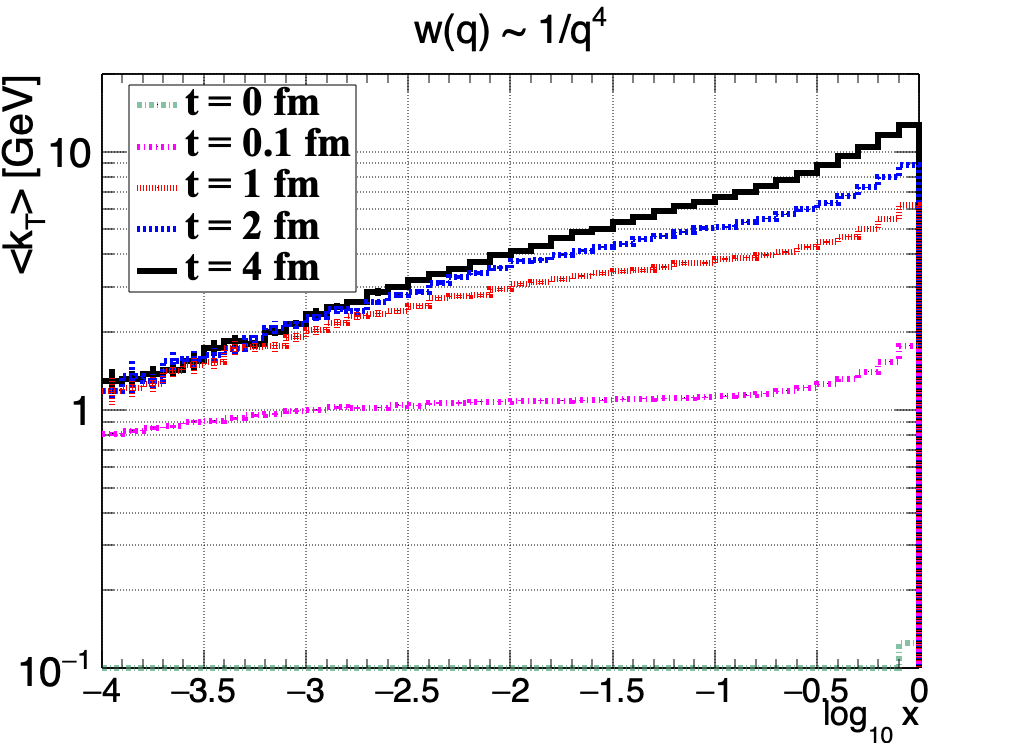} }} 
   \put(800,0){\makebox(0,0)[lb]{\includegraphics[width=80mm, height=60mm]{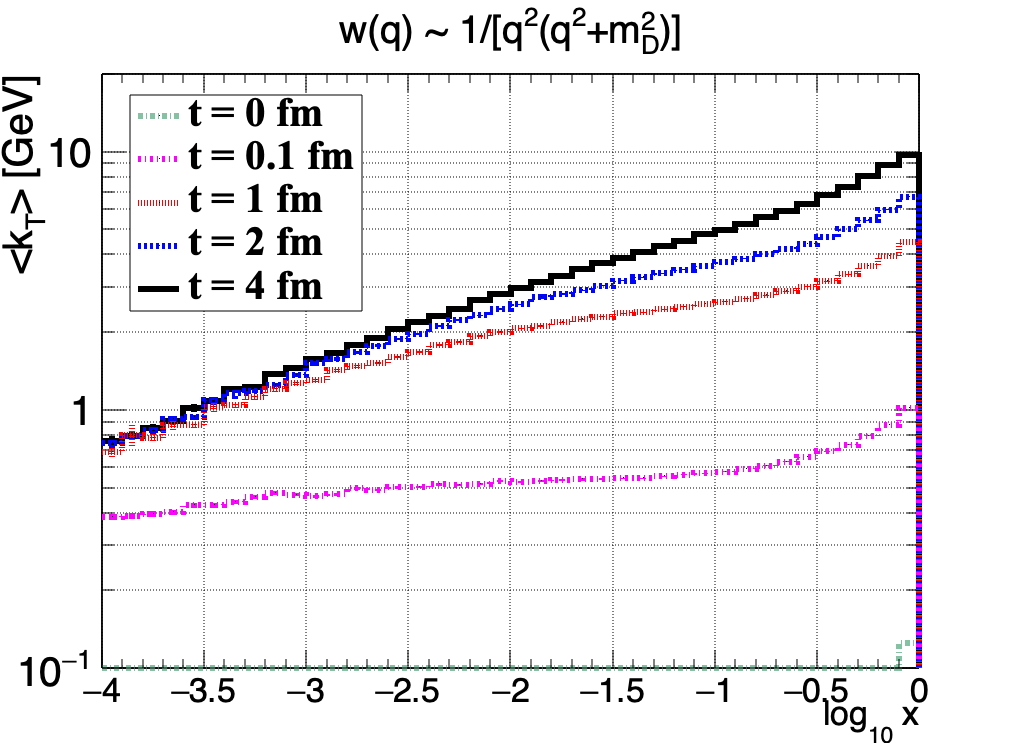} }}
\end{picture}
\caption{\sf 
   The $\langle k_T \rangle$ v.s.\ $\log_{10} x$ distributions from the Monte Carlo program \mincas\ for the evolution time values: $t = 0,\, 0.1,\,1,\,2\,,4\,$fm.
   The LHS figures are for $w(\mathbf{q})$ of Eq.~(\ref{eq:wq1}), while the RHS ones for that of Eq.~(\ref{eq:wq2}).
}
\label{fig:avkTdis}
\end{figure}

In Fig.~\ref{fig:avkTdis} we show the dependence of the $k_T$ mean value  
on the $x$ variable for the $q$-kernel function $w(\mathbf{q})$ 
of Eq.~(\ref{eq:wq1}) (the LHS plot) and for that of Eq.~(\ref{eq:wq2}) 
(the RHS plot).
One can observe increasing $\langle k_T \rangle$ in the course of 
the evolution in the whole $x$ region for small evolution times
and its accumulation in the low-$x$ region for large evolution times.
This pattern is very similar to the one presented in
Ref.~\cite{Blaizot:2014ula} for some approximate analytical solution. 
One should also comment on a distinctive feature of the slope of the final-state cascades $\langle k_T \rangle$ 
as a function of $x$ as compared with the behaviour of $\langle k_T \rangle$ as a function of $x$ 
in the initial-state cascades \cite{Kovchegov:2012mbw}. 
Namely, for the final states one can see that the lower $x$ the typical $k_T$ is lower, 
while for the initial-state cascades the opposite happens. 
From Fig.~\ref{fig:avkTdis} it can also be seen that $\langle k_T \rangle$ rises faster with the evolution time 
for $w(\mathbf{q})$ of Eq.~(\ref{eq:wq1}) than for that of Eq.~(\ref{eq:wq2}).
This can interpreted as a more efficient quenching by the non-equilibrated plasma than by the equilibrated one,
however the shapes are very similar, so the rate of the quenching is similar.  
\begin{figure}[t!]
\setlength{\unitlength}{0.1mm}
  \begin{picture}(1600,1200)
   \put(   0,600){\makebox(0,0)[lb]{\includegraphics[width=80mm, height=60mm]{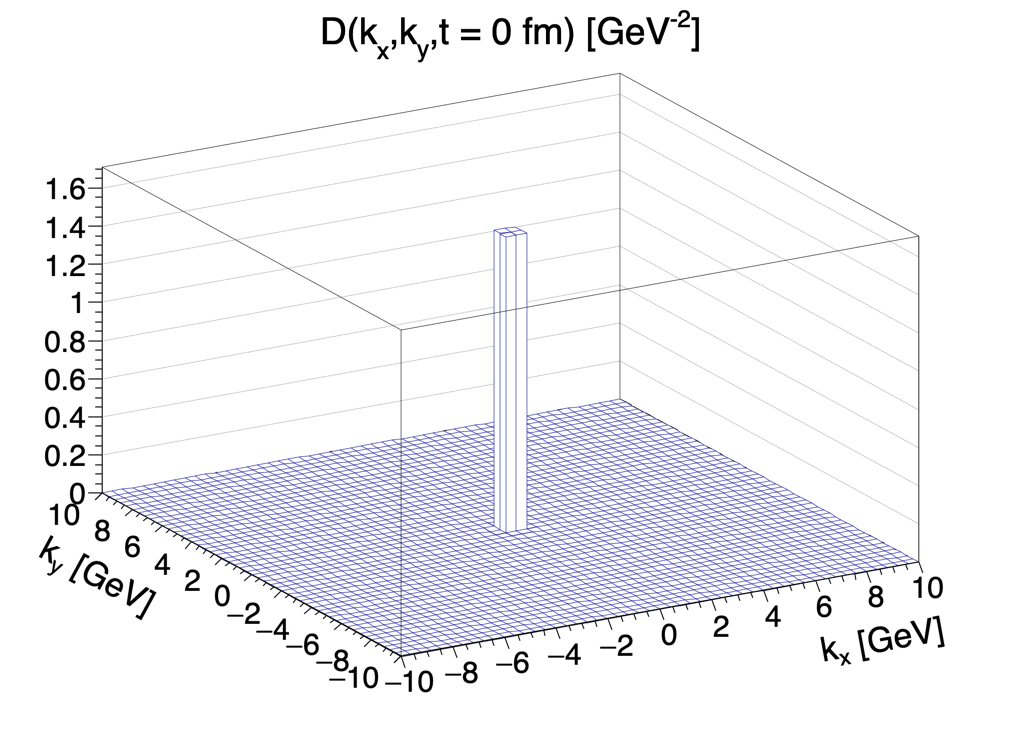} }} 
   \put(800,600){\makebox(0,0)[lb]{\includegraphics[width=80mm, height=60mm]{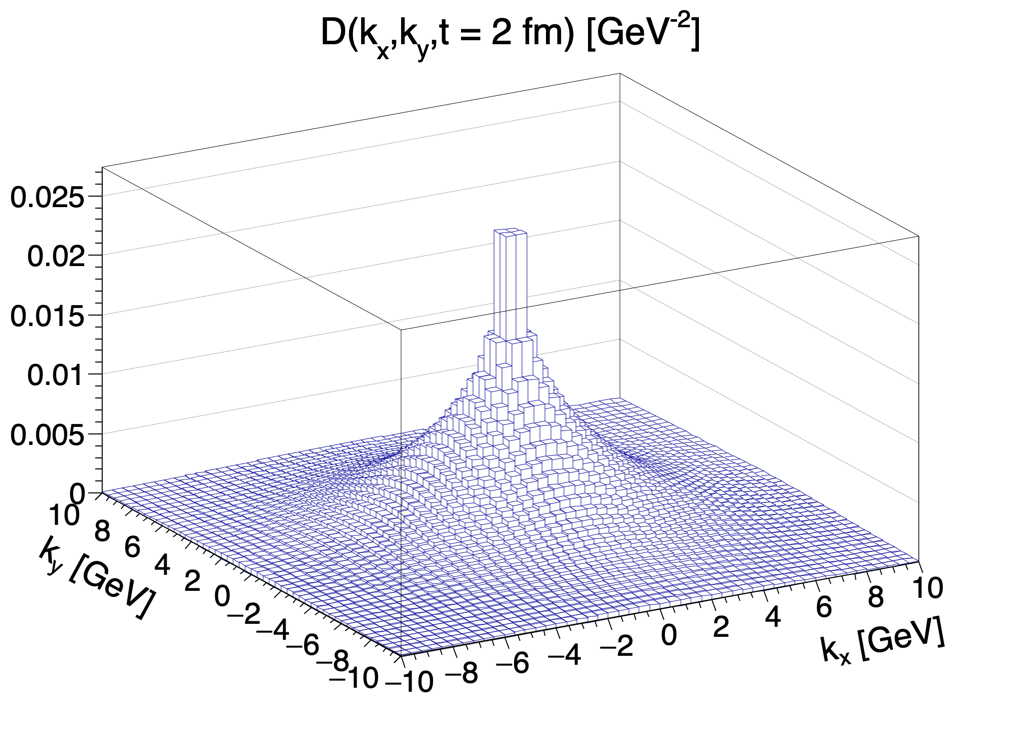} }}
   \put(   0,     0){\makebox(0,0)[lb]{\includegraphics[width=80mm, height=60mm]{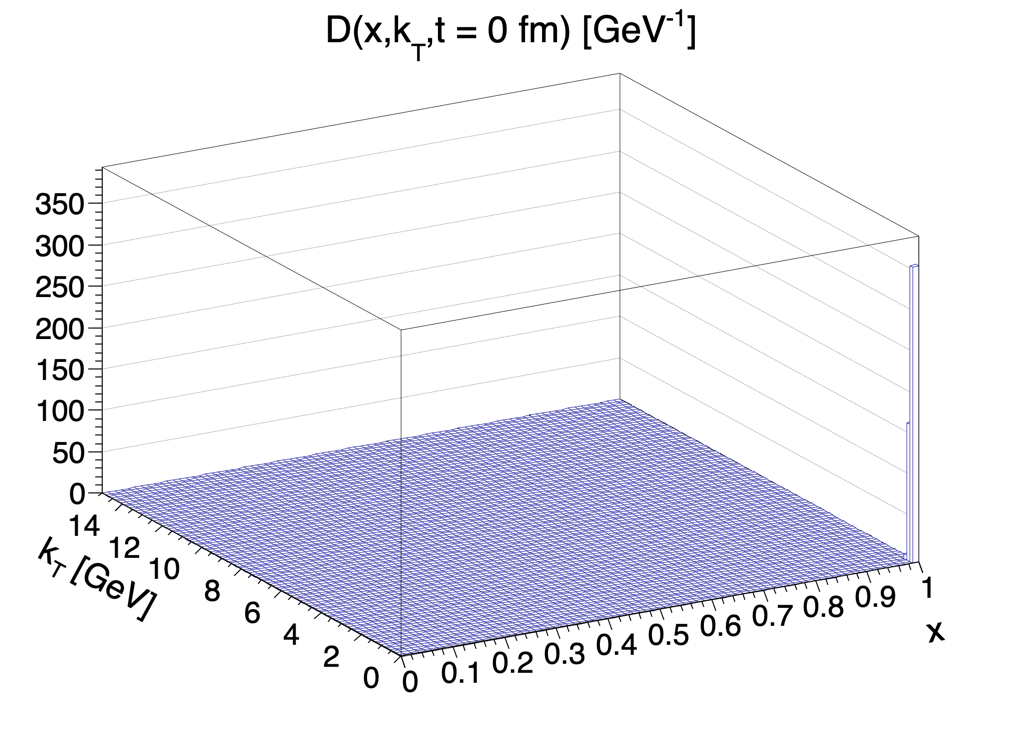} }}
   \put(800,    0){\makebox(0,0)[lb]{\includegraphics[width=80mm, height=60mm]{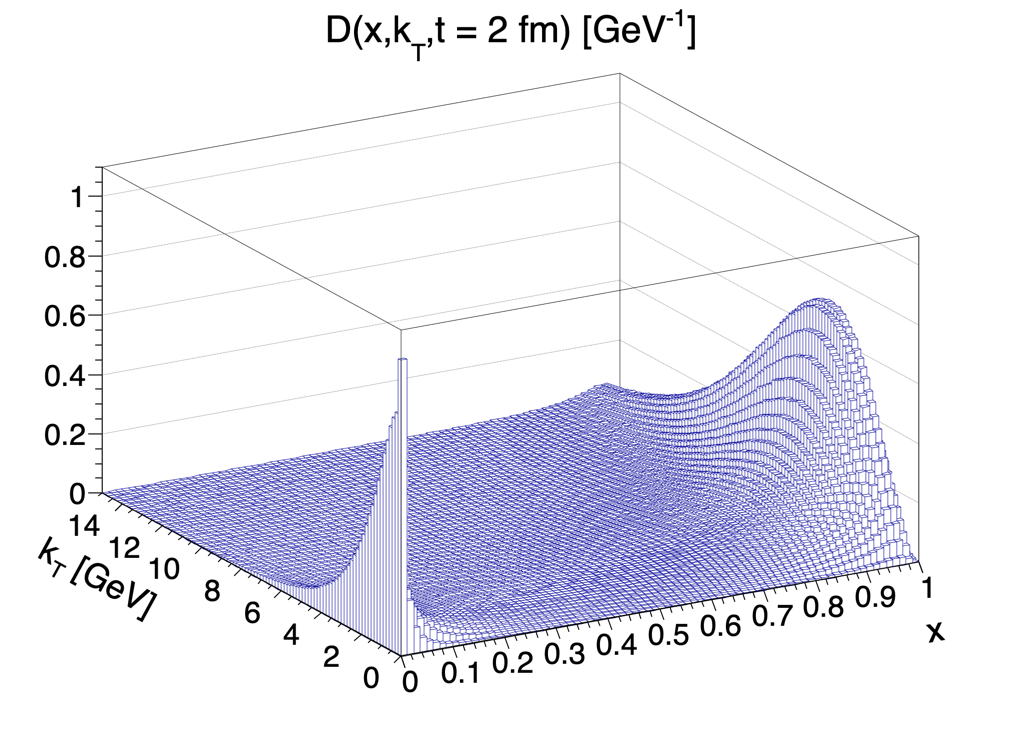} }}
 \end{picture}
\caption{\sf 
   2D distributions of $k_x$ vs.\ $k_y$ (upper row) and  $x$ vs.\ $k_T$ (lower row) from the Monte Carlo program \mincas\ for the evolution time values: $t = 0$ and $2\,$fm (LHS and RHS figures, respectively).
}
\label{fig:kxkyxkT}
\end{figure}

Finally, in Fig.~\ref{fig:kxkyxkT} we show examples of 2D distributions 
of $k_x$ vs.\ $k_y$ (upper row) and  $x$ vs.\ $k_T$ (lower row) 
for the exact $z$ kernel and $w(\mathbf{q})$ of Eq.~(\ref{eq:wq2}). 
The LHS plots present initial distributions, i.e.\ for  $t = 0$, while the RHS ones the evolved distributions at $t = 2\,$fm. 
One can observe how the initial gluon distributions get `diffused' in $x$ and $\mathbf{k}$ in the course of the medium-induced QCD evolution.
The apparent departure from the Gaussian $\mathbf{k}$ distribution can be clearly seen in the upper-right plot.
In the lower plots the turbulent behaviour of the distribution in the $x$ direction, as discussed above,
is also visible.

\section{Summary and outlook}
\label{sec:Sum}

In this paper we have obtained numerical solutions of the equations describing the inclusive gluon distribution 
as produced by a jet the propagating in QGP, given in Ref.~\cite{Blaizot:2014rla}.
These equations were reformulated as the integral equations which allows for their efficient solution using 
the newly constructed Markov Chain Monte Carlo algorithms implemented in the dedicated Monte Carlo program \mincas. 
The results for the energy density (the $x$ distribution) were cross-checked with algorithm based on a direct 
numerical solution of the integro-differential equation by applying the Runge--Kutta-based method, and for the simplified
emission kernel also with the exact analytical solution~\cite{Blaizot:2014rla}.
The MCMC method turns out to be far more efficient in solving the above equations than the differential method. 

The resulting distributions of the gluon density as function of the transverse momenta show some new features, 
not studied so far in the literature on this subject, 
i.e.\ the departure, as the evolution time passes, from the initial Gaussian distribution.
This is a result of the exact treatment of the gluon transverse-momentum broadening due 
to an arbitrary number of the collisions with the medium
together with its shrinking due an arbitrary number of the emission branchings. 
We observe this behaviour for two different forms of the collision kernel $w(\mathbf{q})$.

In the future, we plan to study in a more detailed and systematic way a relation of our MCMC solution 
to the existing approximate solutions as well as to test other possible forms of the collision kernel $w(\mathbf{q})$
and the quenching parameter $\hat q$ resulting from them (in the present study, in order to have 
a correspondence to existing results, we have used the standard value of $\hat q = 1\,$GeV$^2$/fm). 
This will allow to see how universal the pattern of the gluon distribution in QGP is.
For instance, one can use some AdS/CFT models to obtain $w(\mathbf{q})$. 
One can also use our MCMC-based method to solve more general versions of
Eq.~(\ref{eq:ktee1}) or an even more general kinetic equation (which assumes thermalisation of soft gluons) obtained in Ref.~\cite{Baier:2000sb},
and perform a full parton-shower simulation of the final state based on the generated distribution.

\section*{Acknowledgements}
\label{sec:Ack}
We would like to thank Andreas van Hameren, Jacopo Ghiglieri, Bronislav Zakharov for useful comments. 
KK acknowledges the CERN TH Department for hospitality when a large part of this project was done 
and for stimulating discussions with Jacopo Ghiglieri, Ulrich Heinz,  Alexi Kurkela, Konrad Tywoniuk, Urs Wiedemann, 
Bin Wu and Korinna Zapp. 
Furthermore, KK would like to thank Yacine Mehtar-Tani and Jean-Paul Blaizot for informative e-mail exchanges.

\appendix
\section{Branching method with importance sampling}

As was said in Section~\ref{sec:MCMC}, the random variable $z$ cannot
be easily generated according to the {\it pdf} $\zeta(z)$, because it
is a complicated function. 
For this purpose we can utilise the importance sampling 
technique, i.e.\  we can replace $\zeta(z)$ with some
simpler $\tilde{\zeta}(z)$:
\begin{equation}
\begin{aligned}
\tilde{\zeta}(z_i) & = \frac{g(z_i)}{\tilde{\kappa}(\epsilon)}\,, \qquad
g(z_i) =  \frac{1}{(1-z_i)^{3/2}} + \frac{1}{\sqrt{z_i}}\, , \\
\tilde{\kappa}(\epsilon) & = \int_0^{1-\epsilon} dz_i\,g(z_i) 
 = 2\left[\frac{1}{\sqrt{\epsilon}} -1 + \sqrt{1-\epsilon} \right],
\end{aligned}
\label{eq:distzsim}
\end{equation}
and compensate for this simplification with an appropriate 
MC weight. 
The above simplification affects, however, generation of
the random variable $\mathbf{q}_i$ because $z$ has the joint
{\it pdf} with $\mathbf{q}$, namely $\xi(z,\mathbf{q})$ given in
Eq.~(\ref{eq:probzq}), and also generation of $\tau_i$.

In order to describe this in detail, 
let us introduce some useful notation:
\begin{eqnarray}
\tilde{\Psi}(x) &=& \tilde{\Phi}(x) + W\,,
\label{eq:tildPsi} \\
\tilde{\Phi}(x) &=& \frac{1}{\sqrt{x}}\,
                    \tilde{\kappa}(\epsilon)\,,
\label{eq:tildPhi} \\
 \tilde{\varrho}(\tau_i) &=& \tilde{\Psi}(x_{i-1})\, e^{-\tilde{\Psi}(x_{i-1})(\tau_i - \tau_{i-1})}\,,
\label{eq:tildvrho}
\end{eqnarray}
where $W$ is given in Eq.~(\ref{eq:Psi}).
Then, we can express the product of probability densities 
of the random variables $\tau_i$, $z_i$ and $\mathbf{q}_i$ 
in terms of the above functions:
\begin{equation}
\begin{aligned}
~ & \varrho(\tau_i) \xi(z_i,\mathbf{q}_i) = \Psi(x_{i-1}) e^{-\Psi(x_{i-1})(\tau_i - \tau_{i-1})} \, \frac{{\cal G}(z_i,\mathbf{q_i})}{\Psi(x_{i-1}) } \\ 
  & = e^{-\Psi(x_{i-1})(\tau_i - \tau_{i-1})} \left[\frac{z_i{\cal K}(z_i)}{\sqrt{x_{i-1}}}\,\theta(1-\epsilon-z_i) \,\delta(\mathbf{q}_i)  + 
          t^* \, \frac{w(\mathbf{q}_i)}{(2\pi)^2} \,\theta(|\mathbf{q}_i | - q_{\mathrm{min}}) \delta(1-z_i)\right] \\
  & = e^{-\Psi(x_{i-1})(\tau_i - \tau_{i-1})} \bigg[\tilde{\Phi}(x_{i-1})\,\tilde{\zeta}(z_i)\,\frac{z_i{\cal K}(z_i)}{g(z_i)}\,\theta(1-\epsilon-z_i) \,\delta(\mathbf{q}_i) \\
  & \hspace{35mm}  + 
          W\,\omega(\mathbf{q}_i) \,\theta(|\mathbf{q}_i | - q_{\mathrm{min}}) \delta(1-z_i)\bigg] \\
  & = \tilde{\Psi}(x_{i-1}) e^{-\tilde{\Psi}(x_{i-1})(\tau_i - \tau_{i-1})} 
         \bigg[\frac{\tilde{\Phi}(x_{i-1})}{\tilde{\Psi}(x_{i-1})}\,\tilde{\zeta}(z_i)\,\frac{z_i{\cal K}(z_i)}{g(z_i)}\,\theta(1-\epsilon-z_i) \,\delta(\mathbf{q}_i) \\
  & \hspace{30mm}  + 
          \frac{W}{\tilde{\Psi}(x_{i-1})}\,\omega(\mathbf{q}_i) \,\theta(|\mathbf{q}_i | - q_{\mathrm{min}}) \delta(1-z_i)\bigg] \, 
           e^{[{\tilde\Psi}(x_{i-1}) - \Psi(x_{i-1}) ](\tau_i - \tau_{i-1})}\\
  &   = \tilde{\varrho}(\tau_i) 
         \bigg[\,\tilde{p}_i \,\tilde{\zeta}(z_i)\,v(\tau_i,z_i)\,\theta(1-\epsilon-z_i) \,\delta(\mathbf{q}_i) \\ 
 & \hspace{11mm}  
       +   (1-\tilde{p}_i)\,\omega(\mathbf{q}_i)\, h(\tau_i) \,\theta(|\mathbf{q}_i | - q_{\mathrm{min}}) \delta(1-z_i) \bigg],       
\end{aligned}
\label{eq:ktvrhoxi}
\end{equation}
where
\begin{equation}
\tilde{p}_i = 
\frac{\tilde{\Phi}(x_{i-1})}{\tilde{\Psi}(x_{i-1})},\quad 
0\leq  \tilde{p}_i \leq 1\,,
\label{eq:probtil}
\end{equation}
and $v(\tau_i,z_i)$ is the compensating weight for simplifications
done in generation of the random variables $\tau_i$ and $z_i$:
\begin{equation}
v(\tau_i,z_i) = \frac{z_i{\cal K}(z_i)}{g(z_i)}\,e^{[\tilde{\Phi}(x_{i-1})-\Phi(x_{i-1})] (\tau_i-\tau_{i-1})} 
=  \frac{[f(z_i)]^{5/2}}{\sqrt{z_i} + (1-z_i)^{3/2}}\,h(\tau_i)\, ,
\label{eq:wgtz}
\end{equation}
with 
\begin{equation}
h(\tau_i)= e^{\frac{\Delta}{\sqrt{x_{i-1}}}(\tau_i - \tau_{i-1})} \,,
\label{eq:ktwgth}
\end{equation}
where
\begin{equation}
\Delta = \lim_{\epsilon \rightarrow 0} \left[\tilde{\kappa}(\epsilon) - \kappa(\epsilon)\right]
\approx 3.57066164.
\label{eq:diffkap}
\end{equation}
It turns out that for the difference of the integrals $\tilde{\kappa}(\epsilon)$ and $\kappa(\epsilon)$ we
can take the limit $\epsilon \rightarrow 0$ 
-- this limit $\Delta$ is finite and can be computed 
(e.g.\ numerically) once for a given function $f(z)$,
independently of $\epsilon$; 
the above value corresponds to $f(z)$  given in Eq~(\ref{eq:kernel1}) (e.g.~for a simple case of $f(z)=1$: $\Delta = 0$).
This suggests that in the MC generation we can avoid calculation of the complicated $\epsilon$-dependent integral $\kappa(\epsilon)$, 
instead we can replace it with the simple integral $\tilde{\kappa}(\epsilon) $ and compensate for their difference with MC weight of Eq.~(\ref{eq:ktwgth}).

Because of the change in the $\tau$-variable {\it pdf}: 
$\varrho(\tau) \rightarrow \tilde{\varrho}(\tau)$,
we also need to modify accordingly the stopping rule:
\begin{equation}
\begin{aligned}
\int_{\tau}^{\infty} \varrho(\tau_{n+1}) & = e^{-\Psi(x_n) (\tau-\tau_n)} = e^{-\tilde{\Psi}(x_n) (\tau-\tau_n)}\, e^{[\tilde{\Psi}(x_n)-\Psi(x_n)] (\tau-\tau_n)} \\
& = \int_{\tau}^{\infty} \tilde{\varrho}(\tau_{n+1}) \,e^{\frac{\Delta}{\sqrt{x_n}}(\tau-\tau_n)}\,.
\end{aligned}
\label{eq:tildstopr}
\end{equation}
Thus, we can generate $\tau_{n+1}$ according to $\tilde{\varrho}(\tau_{n+1})$ and apply the MC weight
\begin{equation}
s_n(\tau) = e^{\frac{\Delta}{\sqrt{x_n}}(\tau-\tau_n)} .
\label{eq:wgtstop}
\end{equation}

If the initial density $D(x_0,\mathbf{k}_0,\tau_0)$ is 
a complicated function, we can approximate it with some
simpler function $\tilde{D}(x_0,\mathbf{k}_0,\tau_0)$,
construct the corresponding {\it pdf}:
\begin{equation}
\tilde{\eta}(x_0,\mathbf{k}_0) = \frac{\tilde{D}(x_0,\mathbf{k}_0,\tau_0)}{\tilde{d}(\tau_0)},
\qquad 
\tilde{d}(\tau_0) = \int_0^1 dx_0\int d^2 \mathbf{k}_0 \, \tilde{D}(x_0,\mathbf{k}_0,\tau_0),
\label{eq:distx0k0sim}
\end{equation}
and apply the compensating weight
\begin{equation}
u(x_0,\mathbf{k}_0) = \frac{D(x_0,\mathbf{k}_0,\tau_0)}{\tilde{D}(x_0,\mathbf{k}_0,
\tau_0)}.
\label{eq:wgtx0k0}
\end{equation}

Therefore, we can now generate the random variables
$x_0$, $\mathbf{k}_0$, $\tau_i$, $z_i$ and $\mathbf{q}_i$
according to the {\it pdf}s $\tilde{\eta}(x_0,\mathbf{k}_0)$,
$\tilde{\varrho}(\tau_i)$, $\tilde{\zeta}(z_i)$
and $\omega(\mathbf{q}_i)$, respectively, and
to each generated random-walk trajectory $\gamma_n$ of the length $n$ 
apply the MC event-weight
\begin{equation}
\begin{aligned}
n = 0:&\quad
\tilde{w}_{\gamma_0}(x,\mathbf{k},\tau)  = \tilde{d}(\tau_0)\,u(x_0,\mathbf{k}_0) s_0(\tau)\delta(x-x_0)\delta(\mathbf{k}-\mathbf{k}_0),\\
n > 0:&\quad
\tilde{w}_{\gamma_n}(x,\mathbf{k},\tau) = 
\tilde{d}(\tau_0)\, u(x_0,\mathbf{k}_0) 
  \prod_{i=1}^n \left[v(\tau_i,z_i)\theta(\tilde{p}_i - r_i) + h(\tau_i)\theta(r_i - \tilde{p}_i) \right] 
   \\
 &\hspace{25mm} \times
  s_n(\tau)\,\delta(x-x_n)\,\delta(\mathbf{k}-\mathbf{k}_n)\,,
\end{aligned}
\label{eq:ktwgtmc}
\end{equation}
where $r_i\in {\cal U}(0,1)$, i.e.\ it is a random number
from the uniform distribution on $(0,1)$.

One can prove that the expectation value of the above 
weight corresponds to the solution of Eq.~(\ref{eq:ktItfin}),
i.e.
\begin{equation}
E[\tilde{w}_{\gamma}(x,\mathbf{k},\tau)] 
= D(x,\mathbf{k},\tau)\,.
\label{eq:Expwgt}
\end{equation}
In actual MC computations, the above expectation value
is estimated (according to the Law of Large Numbers)
by the arithmetic mean of the event-weight values 
for a given MC sample. Its statistical error is proportional
to $1/\sqrt{N}$, where $N$ is the number of generated MC
events. We skip the proof of Eq.~(\ref{eq:Expwgt}) here
-- it will be given in our future publication devoted to details of the MCMC algorithm. 

In the case of the algorithm for solving Eq.~(\ref{eq:iterfin}),
the above method simplifies to the pure importance sampling
of $z$ without branching into $\mathbf{q}$, 
and $\mathbf{k}_i=\mathbf{0},\,i=0,1,\ldots$.


\begingroup\begin{thebibliography}{10}

\bibitem{Gyulassy:1990ye}
M.~Gyulassy and M.~Plumer, ``{Jet Quenching in Dense Matter}'', {\em Phys.
  Lett.} {\bf B243} (1990)
432--438.

\bibitem{Wang:1991xy}
X.-N. Wang and M.~Gyulassy, ``{Gluon shadowing and jet quenching in A + A
  collisions at s**(1/2) = 200-GeV}'', {\em Phys. Rev. Lett.} {\bf 68} (1992)
1480--1483.

\bibitem{Adler:2002tq}
{STAR} Collaboration, C.~Adler {\em et al.}, ``{Disappearance of back-to-back
  high $p_{T}$ hadron correlations in central Au+Au collisions at
  $\sqrt{s_{NN}}$ = 200-GeV}'', {\em Phys. Rev. Lett.} {\bf 90} (2003) 082302,
\href{http://www.arXiv.org/abs/nucl-ex/0210033}{nucl-ex/0210033}.

\bibitem{Andrews:2018jcm}
H.~A. Andrews {\em et al.}, ``{Novel tools and observables for jet physics in
  heavy-ion collisions}'',
\href{http://www.arXiv.org/abs/1808.03689}{1808.03689}.

\bibitem{Wiedemann:2009sh}
U.~A. Wiedemann, ``{Jet Quenching in Heavy Ion Collisions}'',
  {[Landolt-Bornstein23,521(2010)]},
\href{http://www.arXiv.org/abs/0908.2306}{0908.2306}.

\bibitem{Aad:2010bu}
{ATLAS} Collaboration, G.~Aad {\em et al.}, ``{Observation of a
  Centrality-Dependent Dijet Asymmetry in Lead-Lead Collisions at
  $\sqrt{s_{NN}}=2.77$ TeV with the ATLAS Detector at the LHC}'', {\em Phys.
  Rev. Lett.} {\bf 105} (2010) 252303,
\href{http://www.arXiv.org/abs/1011.6182}{1011.6182}.

\bibitem{Baier:2000mf}
R.~Baier, D.~Schiff, and B.~G. Zakharov, ``{Energy loss in perturbative QCD}'',
  {\em Ann. Rev. Nucl. Part. Sci.} {\bf 50} (2000) 37--69,
\href{http://www.arXiv.org/abs/hep-ph/0002198}{hep-ph/0002198}.

\bibitem{Baier:2000sb}
R.~Baier, A.~H. Mueller, D.~Schiff, and D.~T. Son, ``{'Bottom up'
  thermalization in heavy ion collisions}'', {\em Phys. Lett.} {\bf B502}
  (2001) 51--58,
\href{http://www.arXiv.org/abs/hep-ph/0009237}{hep-ph/0009237}.

\bibitem{Jeon:2003gi}
S.~Jeon and G.~D. Moore, ``{Energy loss of leading partons in a thermal QCD
  medium}'', {\em Phys. Rev.} {\bf C71} (2005) 034901,
\href{http://www.arXiv.org/abs/hep-ph/0309332}{hep-ph/0309332}.

\bibitem{Zakharov:1996fv}
B.~G. Zakharov, ``{Fully quantum treatment of the Landau-Pomeranchuk-Migdal
  effect in QED and QCD}'', {\em JETP Lett.} {\bf 63} (1996) 952--957,
\href{http://www.arXiv.org/abs/hep-ph/9607440}{hep-ph/9607440}.

\bibitem{Zakharov:1997uu}
B.~G. Zakharov, ``{Radiative energy loss of high-energy quarks in finite size
  nuclear matter and quark - gluon plasma}'', {\em JETP Lett.} {\bf 65} (1997)
  615--620,
\href{http://www.arXiv.org/abs/hep-ph/9704255}{hep-ph/9704255}.

\bibitem{Zakharov:1999zk}
B.~G. Zakharov, ``{Transverse spectra of radiation processes in-medium}'', {\em
  JETP Lett.} {\bf 70} (1999) 176--182,
\href{http://www.arXiv.org/abs/hep-ph/9906536}{hep-ph/9906536}.

\bibitem{Baier:1994bd}
R.~Baier, Y.~L. Dokshitzer, S.~Peigne, and D.~Schiff, ``{Induced gluon
  radiation in a QCD medium}'', {\em Phys. Lett.} {\bf B345} (1995) 277--286,
\href{http://www.arXiv.org/abs/hep-ph/9411409}{hep-ph/9411409}.

\bibitem{Baier:1996vi}
R.~Baier, Y.~L. Dokshitzer, A.~H. Mueller, S.~Peigne, and D.~Schiff, ``{The
  Landau-Pomeranchuk-Migdal effect in QED}'', {\em Nucl. Phys.} {\bf B478}
  (1996) 577--597,
\href{http://www.arXiv.org/abs/hep-ph/9604327}{hep-ph/9604327}.

\bibitem{Arnold:2002ja}
P.~B. Arnold, G.~D. Moore, and L.~G. Yaffe, ``{Photon and gluon emission in
  relativistic plasmas}'', {\em JHEP} {\bf 06} (2002) 030,
\href{http://www.arXiv.org/abs/hep-ph/0204343}{hep-ph/0204343}.

\bibitem{Ghiglieri:2015ala}
J.~Ghiglieri, G.~D. Moore, and D.~Teaney, ``{Jet-Medium Interactions at NLO in
  a Weakly-Coupled Quark-Gluon Plasma}'', {\em JHEP} {\bf 03} (2016) 095,
\href{http://www.arXiv.org/abs/1509.07773}{1509.07773}.

\bibitem{Liu:2006ug}
H.~Liu, K.~Rajagopal, and U.~A. Wiedemann, ``{Calculating the jet quenching
  parameter from AdS/CFT}'', {\em Phys. Rev. Lett.} {\bf 97} (2006) 182301,
\href{http://www.arXiv.org/abs/hep-ph/0605178}{hep-ph/0605178}.

\bibitem{Chesler:2014jva}
P.~M. Chesler and K.~Rajagopal, ``{Jet quenching in strongly coupled plasma}'',
  {\em Phys. Rev.} {\bf D90} (2014), no.~2 025033,
\href{http://www.arXiv.org/abs/1402.6756}{1402.6756}.

\bibitem{Carrington:2015xca}
M.~E. Carrington, K.~Deja, and S.~Mrowczynski, ``{Energy Loss in Unstable
  Quark-Gluon Plasma}'', {\em Phys. Rev.} {\bf C92} (2015), no.~4 044914,
\href{http://www.arXiv.org/abs/1506.09082}{1506.09082}.

\bibitem{Mehtar-Tani:2013pia}
Y.~Mehtar-Tani, J.~G. Milhano, and K.~Tywoniuk, ``{Jet physics in heavy-ion
  collisions}'', {\em Int. J. Mod. Phys.} {\bf A28} (2013) 1340013,
\href{http://www.arXiv.org/abs/1302.2579}{1302.2579}.

\bibitem{Ghiglieri:2015zma}
J.~Ghiglieri and D.~Teaney, ``{Parton energy loss and momentum broadening at
  NLO in high temperature QCD plasmas}'', {\em Int. J. Mod. Phys.} {\bf E24}
  (2015), no.~11 1530013, [,271(2016)],
\href{http://www.arXiv.org/abs/1502.03730}{1502.03730}.

\bibitem{Blaizot:2015lma}
J.-P. Blaizot and Y.~Mehtar-Tani, ``{Jet Structure in Heavy Ion Collisions}'',
  {\em Int. J. Mod. Phys.} {\bf E24} (2015), no.~11 1530012,
\href{http://www.arXiv.org/abs/1503.05958}{1503.05958}.

\bibitem{Busza:2018rrf}
W.~Busza, K.~Rajagopal, and W.~van~der Schee, ``{Heavy Ion Collisions: The Big
  Picture, and the Big Questions}'', {\em Ann. Rev. Nucl. Part. Sci.} {\bf 68}
  (2018) 339--376,
\href{http://www.arXiv.org/abs/1802.04801}{1802.04801}.

\bibitem{Salgado:2003gb}
C.~A. Salgado and U.~A. Wiedemann, ``{Calculating quenching weights}'', {\em
  Phys. Rev.} {\bf D68} (2003) 014008,
\href{http://www.arXiv.org/abs/hep-ph/0302184}{hep-ph/0302184}.

\bibitem{Zapp:2008gi}
K.~Zapp, G.~Ingelman, J.~Rathsman, J.~Stachel, and U.~A. Wiedemann, ``{A Monte
  Carlo Model for 'Jet Quenching'}'', {\em Eur. Phys. J.} {\bf C60} (2009)
  617--632,
\href{http://www.arXiv.org/abs/0804.3568}{0804.3568}.

\bibitem{Armesto:2009fj}
N.~Armesto, L.~Cunqueiro, and C.~A. Salgado, ``{Q-PYTHIA: A Medium-modified
  implementation of final state radiation}'', {\em Eur. Phys. J.} {\bf C63}
  (2009) 679--690,
\href{http://www.arXiv.org/abs/0907.1014}{0907.1014}.

\bibitem{Schenke:2009gb}
B.~Schenke, C.~Gale, and S.~Jeon, ``{MARTINI: An Event generator for
  relativistic heavy-ion collisions}'', {\em Phys. Rev.} {\bf C80} (2009)
  054913,
\href{http://www.arXiv.org/abs/0909.2037}{0909.2037}.

\bibitem{Lokhtin:2011qq}
I.~P. Lokhtin, A.~V. Belyaev, and A.~M. Snigirev, ``{Jet quenching pattern at
  LHC in PYQUEN model}'', {\em Eur. Phys. J.} {\bf C71} (2011) 1650,
\href{http://www.arXiv.org/abs/1103.1853}{1103.1853}.

\bibitem{Casalderrey-Solana:2014bpa}
J.~Casalderrey-Solana, D.~C. Gulhan, J.~G. Milhano, D.~Pablos, and
  K.~Rajagopal, ``{A Hybrid Strong/Weak Coupling Approach to Jet Quenching}'',
  {\em JHEP} {\bf 10} (2014) 019, [Erratum: JHEP09,175(2015)],
\href{http://www.arXiv.org/abs/1405.3864}{1405.3864}.

\bibitem{Blaizot:2013vha}
J.-P. Blaizot, F.~Dominguez, E.~Iancu, and Y.~Mehtar-Tani, ``{Probabilistic
  picture for medium-induced jet evolution}'', {\em JHEP} {\bf 06} (2014) 075,
\href{http://www.arXiv.org/abs/1311.5823}{1311.5823}.

\bibitem{Blaizot:2014rla}
J.-P. Blaizot, L.~Fister, and Y.~Mehtar-Tani, ``{Angular distribution of
  medium-induced QCD cascades}'', {\em Nucl. Phys.} {\bf A940} (2015) 67--88,
\href{http://www.arXiv.org/abs/1409.6202}{1409.6202}.

\bibitem{Fister:2014zxa}
L.~Fister and E.~Iancu, ``{Medium-induced jet evolution: wave turbulence and
  energy loss}'', {\em JHEP} {\bf 03} (2015) 082,
\href{http://www.arXiv.org/abs/1409.2010}{1409.2010}.

\bibitem{Blaizot:2014ula}
J.-P. Blaizot, Y.~Mehtar-Tani, and M.~A.~C. Torres, ``{Angular structure of the
  in-medium QCD cascade}'', {\em Phys. Rev. Lett.} {\bf 114} (2015), no.~22
  222002,
\href{http://www.arXiv.org/abs/1407.0326}{1407.0326}.

\bibitem{Iancu:2015uja}
E.~Iancu and B.~Wu, ``{Thermalization of mini-jets in a quark-gluon plasma}'',
  {\em JHEP} {\bf 10} (2015) 155,
\href{http://www.arXiv.org/abs/1506.07871}{1506.07871}.

\bibitem{Gyulassy:1993hr}
M.~Gyulassy and X.-N. Wang, ``{Multiple collisions and induced gluon
  Bremsstrahlung in QCD}'', {\em Nucl. Phys.} {\bf B420} (1994) 583--614,
\href{http://www.arXiv.org/abs/nucl-th/9306003}{nucl-th/9306003}.

\bibitem{Jadach:2002kn}
S.~Jadach, ``{Foam: A General purpose cellular Monte Carlo event generator}'',
  {\em Comput. Phys. Commun.} {\bf 152} (2003) 55--100,
\href{http://www.arXiv.org/abs/physics/0203033}{physics/0203033}.

\bibitem{Galassi:2009}
M.~Galassi {\em et al.}, {\em {GNU Scientific Library Reference Manual}}.
\newblock Network Theory Ltd., 3rd revised edition~ed., 2009.

\bibitem{Christiansen:1970}
J.~Christiansen, ``{Numerical solution of ordinary simultaneous differential
  equations of the 1st order using a method for automatic step change}'', {\em
  Numer. Math.} {\bf 14} (1970) 317--324.

\bibitem{Blaizot:2013hx}
J.-P. Blaizot, E.~Iancu, and Y.~Mehtar-Tani, ``{Medium-induced QCD cascade:
  democratic branching and wave turbulence}'', {\em Phys. Rev. Lett.} {\bf 111}
  (2013) 052001,
\href{http://www.arXiv.org/abs/1301.6102}{1301.6102}.

\bibitem{Kovchegov:2012mbw}
Y.~V. Kovchegov and E.~Levin, {\em {Quantum chromodynamics at high energy}},
  vol.~33.
\newblock Cambridge University Press,
2012.
\newblock

\end{thebibliography}\endgroup
\providecommand{\href}[2]{#2}
\end{document}